# Convolution/deconvolution of generalized Gaussian kernels with applications to proton/photon physics and electron capture of charged particles

W. Ulmer[*]

Klinikum München-Pasing, Department of Radiation Therapy and MPI of Physics, Göttingen, Germany

e-Mail: waldemar.ulmer@gmx.net

**Abstract**

Scatter processes of photons lead to blurring of images produced by CT (computed tomography) or CBCT (cone beam computed tomography) in the KV domain or portal imaging in the MV domain (KV: kilovolt age, MV: megavoltage). Multiple scatter is described by, at least, one Gaussian kernel. In various situations, this approximation is crude, and we need two/three Gaussian kernels to account for the long-range tails (Landau tails), which appear in the Molière scatter of protons, energy straggling and electron capture of charged particles passing through matter and Compton scatter of photons. If image structures are obtained by measurements, these structures are always blurred by scattering. The ideal image (source function) is subjected to Gaussian convolutions to yield a blurred image recorded by a detector array. The inverse problem is to obtain the ideal source image from measured image. Deconvolution methods of linear combinations of two/three Gaussian kernels with different parameters $s_0$, $s_1$, $s_2$ can be derived via an inhomogeneous Fredholm integral equation of second kind (IFIE2) and Liouville - Neumann series (LNS) to provide the source function ρ. A comparison with previously published results is a main purpose in this study. The determination of scatter parameter functions $s_0$, $s_1$, $s_2$ can be best determined by Monte-Carlo simulations. We can verify advantages of the LNS in image processing applied to detector arrays of portal imaging of inverse problems (two/three kernels) of CBCT, IMRT (intensity-modulated radiotherapy), proton scanning beams and IMPT (intensity-modulated proton therapy), where the previous method is partially not applicable. A particular advantage of this procedure is given, if the scatter functions $s_0$, $s_1$, $s_2$ are not constant and depend on coordinates. This fact implies that the scatter functions can be calibrated according to the electron density $ρ_{electron}$ provided by image reconstructions. The convergence criterion of LNS can always be satisfied with regard to the above mentioned cases. A generalization of the present theory is given by an analysis of convolution problems based on the Dirac equation and Fermi-Dirac statistics leading to Landau tails. This generalization is applied to Bethe-Bloch equation (BBE) of charged particles to analyze electron capture. The methodology can readily be extended to other disciplines of physics.

## 1. Introduction

Various scatter processes of photons lead to blurring of images produced by CT/CBCT or portal imaging (KV/MV domain), and similar scatter problems arise in almost all disciplines of physics (e.g. transverse profiles and Bragg curves of proton beams in radiotherapy). Multiple scatter can be described by, at least, one single Gaussian kernel [1 - 6], which we formally abbreviate by K(s, u – x), but it may refer to more than one space dimension. The ideal image (source function ρ without any blurring) is subjected to a Gaussian convolution in order to yield an image function φ (blurred image), which may be recorded by a detector array:

$$K = \frac{1}{s \cdot \sqrt{\pi}} \cdot \exp(-(u-x)^2/s^2)$$
$$\varphi = \int K(s, u-x) \cdot \rho(u) du \quad (1)$$

The magnitude of the parameter s represents a measure of the severeness of blurring that as s → 0 the kernel K tends to the δ-distribution and φ becomes identical with ρ.

In many situations the restriction to one Gaussian kernel represents a crude approximation, and we need a linear combination of Gaussian kernels with $K_g$ as a resulting convolution kernel to account for long-range tails, which appear in the Molière multiple scatter theory of protons and inclusion of Landau tails [1-6, 8 - 9, and references therein] or in Compton scatter of γ-quanta [7, and references therein]:

$$K_g = c_0 \cdot K_0 + c_1 \cdot K_1 + c_2 \cdot K_2. \quad (2)$$
$$K_j = \frac{1}{s_j \cdot \sqrt{\pi}} \cdot \exp(-(u-x)^2/s_j^2) \quad (j = 0, 1, 2). \quad (2a)$$
$$\varphi = \int K_g(s_0, s_1, s_2, c_0, c_1, c_2, u-x) \cdot \rho(u) du. \quad (2b)$$
$$c_0 + c_1 + c_2 = 1. \quad (2c)$$

In every case, the parameters in equation (2c) have to satisfy $c_0 > c_1$, $c_0 > c_2$ and $s_0 < s_1$, $s_0 < s_2$. The restriction to two Gaussian kernels results by setting $c_2 = 0$. If $c_2 \neq 0$, $c_1$ can also be less than zero, but $K_g \geq 0$ has still to be valid. The previously published method [5] of the inverse task of $K_g$ requires $s_1 \wedge s_2 > s_0 \cdot \sqrt{2}$. This restriction can lead to critical cases (proton dosimetry, image processing with CBCT). Therefore the LNS method can circumvent this restriction, since it only needs that $s_1 \wedge s_2 > s_0$ is satisfied (section 2.2).

The inverse problem of equation (1) is to determine the ideal source image from a really determined image. If the scatter parameters are known (e.g. *rms* value s of Gaussian kernels via appropriate test measurements or Monte-Carlo simulations), we are able to calculate the idealistic source structure by an inverse kernel $K^{-1}(s, u - x)$:

$$\rho(x) = \int K^{-1}(s, u-x) \cdot \varphi(u) du. \quad (3)$$

Due to many applications the inverse kernel $K^{-1}(s, u-x)$ of a single Gaussian kernel $K(s, u - x)$ is a proven tool circumventing ill-posed aspects [9 – 13, 22 – 24, 26, 27]. A possible representation of the inverse kernel $K^{-1}(s, u - x)$ is given by:

$$K^{-1}(s, u-x) = \left. \sum_{n=0}^{N} c_n(s) \cdot H_{2n}\left(\frac{u-x}{s}\right) \cdot K(s, u-x) \right\} \quad (4)$$
$$N \to \infty; \quad c_n = (-1)^n \cdot s^{2n} / (2^n \cdot n!); \quad (n = 0, 1, ...., \infty)$$

$H_{2n}$ refer to Hermite polynomials of even order and the inverse kernel $K^{-1}$ can be regarded as a generalized Gaussian convolution kernel. The coefficients $c_n$ of the two-point Hermite polynomials of $K^{-1}$ are determined by a Lie series expansion. Both kernels K and $K^{-1}$ shall be derived as Green's functions in the following section. For practical applications, we have to restrict N to a finite limit (N < ∞), and the question arises, which N provides sufficiently accurate results. Based on formula (4) there have been put forward many applications in radiation physics, mainly with regard to the deconvolution problem of the finite detector size in radiation profiles [9 – 13]. It should also be noted that the simplest, but well-known solution function $c(\zeta, t) = N(t) \cdot \exp(-\zeta^2/(4Dt))$ of the heat/diffusion equation is a Gaussian kernel [14 – 18]. The initial condition of this solution function implies a δ-function resulting from $c(\zeta, t \to 0) = \delta(\zeta)$. The inverse problem of this distribution function [16 – 18] is similar to the problem given by equation (3); it represents a typical case of an ill-posed problem and requires regularizations techniques (the inverse of the δ-function is not defined), which have been studied by the aforementioned authors. However, it appears that in this field the EM algorithm [19 – 21] has proven to be valuable.

The intention of this study is to extend these considerations to the inverse problem of a linear combination of two/three Gaussian convolution kernels $K_g^{-1}(s_0, c_0, s_1, c_1, s_2, c_2, u - x)$ according to equations (2 – 2c), in order to found applications to aforementioned image processing, where a single Gaussian kernel would represent a crude approximation. The kernels $K_g(s_0, c_0, s_1, c_1, s_2, c_2, u - x)$ and $K_g^{-1}$ account for long-range tails in multiple scatter problems such as Landau tails. Since the resulting kernels $K_g$ incorporate linear combinations with different *rms*-values $s_0, s_1, s_2$, they may be regarded as Gaussian-like with long-range tails as being requested in many tasks. In this communication, we shall develop a new solution procedure of the inverse problem of a linear combination of two/three Gaussian kernels, which avoids the determination of the deconvolution kernel $K_g^{-1}$, namely its formulation by an IFIE2 and related LNS to calculate solutions in every desired order. The results obtained by the LNS procedure will be compared with the different procedure to calculate $K_g^{-1}$ from $K_g$, which has been previously published. With regard to applications we preferably consider problems of image processing in the KV and MV domain. In the next section, the inverse kernel $K_g^{-1}$ will be developed according to an IFIE2 and LNS procedure; it represents a tool in IMRT and intensity-modulated proton therapy (IMPT), see e.g. [8, 22 - 24]. We should also point out that in many problems of deconvolutions fast Fourier transforms (FFT) together with Wiener filters are applied. A very concise paper on Fourier-based deconvolutions and filter functions has been given in a review paper [25]. However, some critical aspects result from Fourier-based deconvolutions applied to step functions and are usually referred to as 'ill-posed' (see some applications given in section 3). These well-known problems have already accounted for [5, 23 – 26, 29]. Since Gaussian and Gaussian-like convolutions/deconvolutions play a significant role in many disciplines of physics, engineering [5 – 13, 19 – 29], electron capture of charged particles by passing though matter (based on the Dirac equation and Fermi-Dirac statistics this aspect will be discussed in section 2.6), and even in spectroscopic tasks in molecular biology (e.g. removal of scatter in structure elucidations of bio-molecules by nuclear magnetic resonance (NMR) and X-rays), reliable toolkits for inverse procedures are desirable, which are able to circumvent ill-posed aspects under some restrictions. It should be pointed out that further various applications of Gaussian convolutions/deconvolutions with regard to statistical problems in disciplines beyond the physical scope which we did not discuss here can easily be obtained by a look at internet.

## 2. Methods

In many problems of mathematical physics it is convenient to start with a differential operator formulation and, thereafter, to pass to the corresponding integral equation via Green's function method. Thus the path integral quantization is a very prominent example [32]. At first, we shortly summarize previous results [5, 9], which should be consulted by those readers with need of more detailed information. A very convenient way is the operator formulation to derive the Gaussian convolution kernel as a Green's function and the related inverse problem.

### *2.1. Operator calculus (Lie series of operators) and the derivation of the inverse kernels*

The basic formulas of all subsequent procedures and calculations are the following two operator functions:

$$O = \exp(-\frac{1}{4} \cdot s^2 \cdot \frac{d^2}{dx^2}) . \qquad (5)$$

$$O^{-1} = \exp(\frac{1}{4} \cdot s^2 \cdot \frac{d^2}{dx^2}) . \qquad (5a)$$

The operators O and $O^{-1}$ and their related actions to a class of functions are formally defined by Taylor series of the exponential functions, which represent Lie series of operator functions [5 – 7, 15, 32 – 33]:

$$O^{-1} = 1 + \sum_{n=1}^{\infty} \frac{s^{2n}}{n! \cdot 4^n} \cdot d^{2n}/dx^{2n} \ . \tag{6}$$

$$O = 1 + \sum_{n=1}^{\infty} \frac{s^{2n}}{n! \cdot 4^n} \cdot (-1)^n \cdot d^{2n}/dx^{2n} \ . \tag{6a}$$

O and $O^{-1}$ obey the following relation:

$$\left.\begin{array}{l} O \cdot O^{-1} = O^{-1} \cdot O = \exp(-\frac{1}{4} \cdot s^2 \cdot \frac{d^2}{dx^2}) \cdot \exp(\frac{1}{4} \cdot s^2 \cdot \frac{d^2}{dx^2}) \\ = \exp\left(-\frac{1}{4} \cdot s^2 \cdot \frac{d^2}{dx^2} + \frac{1}{4} \cdot s^2 \cdot \frac{d^2}{dx^2}\right) = 1 \ (unit\ operator) \end{array}\right\}. \tag{7}$$

In three dimensions, we have to substitute operator $d^2/dx^2$ by the 3D Laplace operator $\Delta$:

$$O = \exp(-\frac{1}{4} \cdot s^2 \cdot \Delta) \ . \tag{8}$$

$$O^{-1} = \exp(\frac{1}{4} \cdot s^2 \cdot \Delta). \tag{8a}$$

If $\rho(x)$ represents a source and $\varphi(x)$ an image function, the following relationships have to be satisfied:

$$\varphi(x) = O^{-1} \cdot \rho(x). \tag{9}$$
$$\rho(x) = O \cdot \varphi(x). \tag{9a}$$

It can be concluded from relations (6 – 9a) that all permitted functions $\rho(x)$ and $\varphi(x)$ have to belong to the space $C^\infty$ (Banach space), which implies that both sets $\varphi(x)$ and $\rho(x)$ are defined by derivatives of infinite order. According to the relations (9) and (9a) the integral operator notations of equations (1) and (3) have to represent Green's functions of $O^{-1}$ and $O$:

$$\varphi = \int K(s, u-x) \cdot \rho(u) du \ . \tag{10}$$
$$\rho = \int K^{-1}(s, u-x) \cdot \varphi(u) du \ . \tag{10a}$$

The integral operator kernel $K(s, u-x)$ is the normalized Gaussian kernel of equation (1), which may be based on the spectral theorem of functional analysis [5]. The essential difference between differential and integral operator formulation is the class of the permitted functions.

The differential operator formulations according to relations (9 and 9a) require the only restriction that $\varphi$ and $\rho$ belong to $C^\infty$. By that, the action of the operator $O$ does not lead to an ill-posed operation and to a necessary regularization, as long as the requirement of a norm has not to be account for with regard to $\varphi$ and $\rho$.

6The following relations are valid for all kinds of Green's function, i.e. a jump at u = x:

$$O \cdot K(s, u-x) = \delta(u-x). \tag{11}$$

$$O^{-1} \cdot K^{-1}(s, u-x) = \delta(u-x). \tag{11a}$$

According to the rules of Lie series the multiplication of O·K with $O^{-1}$ from the left-hand side implies the expression K(s, u - x) = $O^{-1}$·δ(u - x).

With the help of the Fourier representation of δ(u-x) the operation $O^{-1}$·δ(u-x) provides:

$$K = O^{-1} \cdot \delta(u-x) = \frac{1}{2\pi} \int O^{-1} \cdot \exp(i \cdot k \cdot u) \cdot \exp(-i \cdot k \cdot x) dk$$

$$= \frac{1}{2\pi} \int \exp(i \cdot k \cdot u) \cdot \exp(-i \cdot k \cdot x) \cdot \exp(-0.25 \cdot s^2 \cdot k^2) dk. \tag{12}$$

The evaluation of equation (12) provides the Gaussian kernel (1).

We now perform the identical procedure via multiplication of equation (11a) with the operator O. By that we obtain:

$$\left. \begin{array}{l} O \cdot O^{-1} \cdot K^{-1}(s, u-x) = K^{-1}(s, u-x) = O \cdot \delta(u-x) = \frac{1}{2\pi} \int O \cdot \exp(i \cdot k \cdot u) \cdot \exp(-i \cdot k \cdot x) dk \Rightarrow \\ \exp(\varepsilon \cdot \frac{s^2}{4} \cdot d^2/dx^2) \cdot \frac{1}{2\pi} \cdot \int dk \cdot \exp(ik(u-x)) = \frac{1}{2\pi} \int \exp(i \cdot k \cdot (u-x)) \cdot \exp(-\varepsilon \cdot \frac{1}{4} \cdot s^2 \cdot k^2) dk. \\ \varepsilon = -1 \end{array} \right\} \tag{12a}$$

The Fourier transform of δ(u-x) in the right-hand side of equation (12a) leads to the term $\exp(0.25 \cdot k^2 \cdot s^2)$ and, the inverse kernel $K^{-1}$ assumes an awkward feature. The source function ρ may be determined by evaluation of the following integral:

$$\rho(x) = \frac{1}{2\pi} \int \varphi(u) \int \exp(0.25 \cdot s^2 \cdot k^2) \cdot \exp(i \cdot k \cdot u) \cdot \exp(-i \cdot k \cdot x) dk du .. \tag{12b}$$

Thus $K^{-1}$(s, u – x) can only be regularized, if *φ(u)* vanishes sufficiently fast, and the Green's function related to the operator O cannot be derived from the analogue expression of formula (12a). In order to derive the integral operator kernel $K^{-1}$ of the operator O via Lie series, we have to carry out some operations in equations (11) and (11a):

$$O^2 \cdot K(s, u - x) = O \cdot \delta(u - x) = K^{-1}(s, u - x). \tag{13}$$

$$O \cdot O^{-1} \cdot K^{-1}(s, u - x) = K^{-1}(s, u - x) = O \cdot \delta(u - x). \tag{13a}$$

By elimination of O·δ(u − x) in both equations (13) and (13a) the inverse operator $K^{-1}$ can be constructed. The most essential feature results from the operator $O^2$:

$$O^2 = \exp(-0.5 \cdot s^2 \cdot d^2/dx^2). \tag{13b}$$

The action of the operator $O^2$ is obtained by its Lie series applied to K:

$$K^1(s, u-x) = O^2 \cdot K(s, u-x) = \sum_{n=0}^{\infty} (-1)^n \cdot 2^{-n} \cdot \frac{1}{n!} \cdot d^{2n}/dx^{2n} \cdot K(s, u-x). \tag{13c}$$

The right-hand side of equation (13c) can be written in terms of Hermite polynomials, which yields relation (4). With the help of relations (10a) and (13c) we are able to calculate the inverse problem of a Gaussian convolution by two different ways; both result from equation (13d):

$$\rho(x) = \int K^{-1}(s, u-x)\varphi(u)du = O^2 \cdot \int K(s, u-x)\varphi(u)du = \sum_{n=0}^{\infty} (-1)^n \cdot 2^{-n} \cdot \frac{1}{n!} \cdot s^{2n} \int H_{2n}(\frac{u-x}{s}) \cdot K(s, u-x)\varphi(u)du. \tag{13d}$$

1. Integration with the kernel K (convolution) and subsequent differentiation of the result with $O^2$.
2. Integration of φ with the kernel $K^{-1}$, i.e. the Hermite polynomials are accounted for in all terms.

As already mentioned the integral operator K only requires the Banach space $L_1(-\infty, +\infty)$ of Lebesque-integrable functions, whereas with regard to $K^{-1}$ there are some restrictions. However, in those cases, where φ is only non-vanishing in a finite interval with $L_1(a, b)$, the inverse problem with the integral operator $K^{-1}$ always exists. This fact has an important meaning in practical applications, where summations in finite intervals have to be accounted for (step functions, voxel integrations). The integral operator correspondence to the relation $O \cdot O^{-1} = 1$ is given by the following equation:

$$\int K(s, u - u') \cdot K^{-1}(s, u' - x) du' = \delta(u - x). \tag{14}$$

It has to be mentioned that relation (14) is also valid for every kind of integral operators, if both K and $K^{-1}$ exist. There are various problems, where the differential operator calculus is easier to handle, e.g. the derivation of basic formulas, and we mention the properties of iterated kernels. The repeated application (*n times*) of the operators $O^{-1}$ and $O$ implies the expressions:

$$\left.\begin{aligned} O^{-1} \cdot \ldots \cdot O^{-1} \, (n \text{ times}) &= O^{-n} = \exp\left(\frac{s_n^2}{4} \cdot \frac{d^2}{dx^2}\right) \\ O \cdot \ldots \cdot O \, (n \text{ times}) &= O^n = \exp\left(-\frac{s_n^2}{4} \cdot \frac{d^2}{dx^2}\right) \\ s_n^2 &= n \cdot s^2 \end{aligned}\right\} \quad (15)$$

The integral operator kernels of relation (15) are simply given by the modification:

$$\left.\begin{aligned} K(s, u-x) &\Rightarrow K(s_n, u-x); \; K^{-1}(s, u-x) \Rightarrow K^{-1}(s_n, u-x) \\ s^2 &\Rightarrow s_n^2 = n \cdot s^2 \end{aligned}\right\} \quad (16)$$

The 3D extension of the relation (1) is the 3D Gaussian convolution kernel, which reads:

$$K(s, u-x, v-y, w-z) = \frac{1}{\sqrt{\pi}^3} \cdot \frac{1}{s^3} \cdot \exp\left(-\frac{1}{s^2} \cdot ((u-x)^2 + (v-y)^2 + (w-z)^2)\right). \quad (17)$$

Integrations have to be carried out over $u$, $v$ and $w$. The integral operator correspondence $K^{-1}$ of equation (17) is obtained in a similar way. For this purpose, we write the Hermite polynomial expansion in each dimension according to equation (4) by introducing the terms $F_1(s, u-x)$, $F_2(s, v-y)$ and $F_3(s, w-z)$. $F_1$, $F_2$ and $F_3$ will be defined as below and have to be multiplied with the kernel K according to equation (17). By that, we obtain:

$$\left.\begin{aligned} F_1(s, u-x) &= \sum_{n=0}^{N} c_n(s) \cdot H_{2n}(\tfrac{u-x}{s}); \; F_2(s, v-y) = \sum_{n=0}^{N} c_n(s) \cdot H_{2n}(\tfrac{v-y}{s}); \; F_3(s, w-z) = \sum_{n=0}^{N} c_n(s) \cdot H_{2n}(\tfrac{w-z}{s}) \\ N &\to \infty \end{aligned}\right\} \quad (18)$$

$$K^{-1}(s, u-x, v-y, w-z) = \prod_{k=1}^{3} F_k \cdot K(s, u-x, v-y, w-z). \quad (19)$$

Equation (19) is also valid, if the substitution $s^2 \to s_n^2 = n \cdot s^2$ is performed. In the preceding section we have stated arguments, why in some situations a linear combination of Gaussian convolution kernels is required according to equation (2). An important feature of the operators $O$ and $O^{-1}$ and the related integral operator notation is the class of functions, for which they are defined.

1. *Both operators $O$ and $O^{-1}$ act on the set of functions, which belong to $C^\infty$.*

This is valid even for functions like $\exp(x^2)$, etc. In contrast to these differential operators, the kernels K and $K^{-1}$ may be associated with a norm (exception: the subsequent point 2). In QM the situation is that the Schrödinger differential equation formulation corresponds to $O$ and/or $O^{-1}$, whereas the Green's function approach is the Feynman path integral quantization [32]. The common restriction in QM is the normalization condition of the Hilbert space [32, 34], which selects the set of permitted functions in both cases. Nevertheless, for scatter problems the path integrals provide more flexibility.

2. *The function class of finite polynomials $x^n$ and linear combinations $(a_0 + a_1 \cdot x + ...., + a_n \cdot x^n)$ belong to $C^\infty(-\infty, \infty)$, but not to $L_1(-\infty, \infty)$ and $L_2(-\infty, \infty)$.*

1.

The integral operators K and $K^{-1}$ lead to the identical results as O and $O^{-1}$, and it is easy to verify that the operations belong to $C^\infty$. The integrals (22) can be evaluated with binominal theorem. In equation (22a) $O^2$ acts on the substitution $(s \cdot \zeta + x)^n$; this procedure is easier to handle than to use Hermite polynomials (this might be intricate).

2.

$$O^{-1} \cdot x^n = O^{-1} \cdot x^n = x^n + \frac{s^2}{4} \cdot n \cdot (n-1) \cdot x^{n-2} + ... + \frac{s^{2m}}{4^m \cdot m!} \cdot n \cdot (n-1) \cdot ... \cdot (n - 2m + 1) \cdot x^{n-2m} \quad (20)$$
$$2m \leq n$$

$$O \cdot x^n = x^n - \frac{s^2}{4} \cdot n \cdot (n-1) \cdot x^{n-2} + ... - ... + (-1)^m \frac{s^{2m}}{4^m \cdot m!} \cdot n \cdot (n-1) \cdot ... \cdot (n - 2m + 1) \cdot x^{n-2m} \quad (21)$$
$$2m \leq n$$

$$\int K(s, u - x) \cdot u^n \, du = \int K(\zeta) \cdot (s \cdot \zeta + x)^n \cdot \frac{1}{s} \cdot d\zeta \quad (22)$$
$$\zeta = \frac{u-x}{s}$$

$$\int K^{-1}(s, u - x) \cdot u^n \, du = O^2 \cdot \int K(\zeta) \cdot (s \cdot \zeta + x)^n \cdot \frac{1}{s} \cdot d\zeta. \quad (22a)$$

The function f(x) is Lesbesque-measureable in a finite interval $[a \leq x \leq b]$ and vanishes outside. Then the integrations over K·f and $K^{-1}$·f exist in $L_1$ and $L_2$, whereas the differentiations according to the properties of O and $O^{-1}$ require smooth functions f(x) of the class $C^\infty(a, b)$. Measurement data may be refined by spline functions. This problem occurs with regard to back convolutions of measurement data given in digital form.

Now we finally wish discuss the problem of the permitted class of functions with respect to two prominent examples associated with the kernels $K^{-1}$ and K in $L_1(-\infty, \infty)$ and $L_2(-\infty, \infty)$.

Keeping the relations (19 – 22a) in mind, we consider g(x) as Gaussian function multiplied with Hermite polynomials as test functions. It is known that this class is complete in the Banach space $L_1$ and Hilbert space $L_2$ [34 - 35]. In order to show the calculation procedure, we restrict ourselves at first to a single Gaussian as test function subjected to a Gaussian convolution/deconvolution. Thus we can solve this task by two different ways:

$$g(x) = \frac{1}{\sigma \cdot \sqrt{\pi}} \cdot \exp(-x^2/\sigma^2)$$

$$\hat{g}(x) = O^{-1} \cdot \exp(-x^2/\sigma^2) \cdot \frac{1}{\sigma \cdot \sqrt{\pi}} = \exp(\tfrac{1}{4} \cdot s^2 \cdot \tfrac{d^2}{dx^2}) \cdot \exp(-x^2/\sigma^2) \cdot \frac{1}{\sigma \cdot \sqrt{\pi}} \quad (23)$$

$$\hat{g}(x) = \int_{-\infty}^{\infty} K(s, u-x) \cdot \frac{1}{\sigma \cdot \sqrt{\pi}} \cdot \exp(-u^2/\sigma^2) du = \frac{1}{\sqrt{\pi} \cdot \sqrt{\sigma^2 + s^2}} \cdot \exp(-x^2/(\sigma^2 + s^2)). \quad (23a)$$

It should be pointed out that the determination of ĝ(x) via differential operator expansion requires a lot of effort, whereas the calculation of ĝ(x) via usual convolution only needs the substitutions:

$$\xi = \tfrac{1}{s} \cdot \sqrt{\sigma^2 + \varepsilon \cdot s^2} \cdot \tfrac{1}{\sigma} \cdot u - \frac{1}{\sqrt{\sigma^2 + \varepsilon \cdot s^2}} \cdot \sigma \cdot x. \quad (24)$$

$$du = \frac{s \cdot \sigma}{\sqrt{\sigma^2 + \varepsilon \cdot s^2}} \cdot d\xi; \quad (\varepsilon = 1). \quad (24a)$$

We can represent every Hermite polynomial $H_n(\xi)$ in terms of ordinary polynomials by the following formulas [7, 35]:

$$\left.\begin{array}{l} H_n(\xi) = 2^n \cdot \xi^n + n \cdot \sum_{i=1}^{I} \left((-1)^i \cdot \prod_{j=1}^{i}(n-j) \cdot 2^{n-2i} \cdot \xi^{n-2i}\right) \quad (I=(n-1)/2; \; n=1,3,5,7,\ldots) \\ H_{n+1}(\xi) = 2 \cdot \xi \cdot H_n(\xi) - dH_n(\xi)/d\xi \quad (n=0,1,2,3,\ldots) \\ H_0 = 1 \end{array}\right\} \quad (25)$$

Therefore it is sufficient to consider for the more general case the set of test functions:

$$g(x) = \frac{1}{\sigma \cdot \sqrt{\pi}} \cdot \exp(-x^2/\sigma^2) \cdot x^n. \quad (26)$$

The above substitution formulas (24 and 24a) remain unchanged, but the expression $u^n$ has to be calculated according to

$$u^n = (s \cdot \xi + \sigma \cdot \frac{1}{\sqrt{\sigma^2 + \varepsilon \cdot s^2}} \cdot x)^n \cdot \sigma^n \cdot \frac{1}{(\sqrt{\sigma^2 + \varepsilon \cdot s^2})^n}. \quad (27)$$

The evaluation of the following integral requires the binominal theorem:

$$\left.\begin{array}{l} u^n = \alpha_n \cdot \sum_{j=0}^{n} \binom{n}{j} \cdot s^j \cdot \xi^j \cdot \beta^{n-j} \cdot x^{n-j} \\ \alpha_n = \sigma^n \cdot (\sigma^2 + \varepsilon \cdot s^2)^{-n/2}; \quad \beta = \sigma \cdot (\sigma^2 + \varepsilon \cdot s^2)^{-1/2} \end{array}\right\} \quad (28)$$

By that, we obtain:

$$\left.\begin{aligned}\hat{g}(\sigma, s, \varepsilon, x) &= \int_{-\infty}^{\infty} K(s, u-x) \cdot \frac{1}{\sigma \cdot \sqrt{\pi}} \cdot u^n \cdot \exp(-u^2/\sigma^2) du \\ \rightarrow \hat{g}(\sigma, s, \varepsilon, x) &= \int_{-\infty}^{\infty} K(s, u-x) \cdot \frac{1}{\sigma \cdot \sqrt{\pi}} \cdot \alpha_n \cdot \sum_{j=0}^{n} \binom{n}{j} \cdot s^j \cdot \xi^j \cdot \beta^{n-j} \cdot x^{n-j} \cdot \exp(-u^2/\sigma^2) du\end{aligned}\right\}. \quad (29)$$

We use again the above substitutions (24, 24a) then the above integral can be solved with the Γ-function:

$$\int_{-\infty}^{\infty} \xi^j \cdot \exp(-\xi^2) d\xi = 2 \cdot \int_{0}^{\infty} (1*(-1)^j) \cdot 0.5 \cdot \xi^j \cdot \exp(-\xi^2) d\xi = \tfrac{1}{2} \cdot (1+(-1)^j) \cdot \Gamma(\tfrac{j+1}{2}). \quad (30)$$

The complete solution of equation (29) can be written as:

$$\left.\begin{aligned}\hat{g} &= \int_{-\infty}^{\infty} u^n \cdot \frac{1}{\sigma \cdot \sqrt{\pi}} \cdot \exp(-u^2/\sigma^2) \cdot \exp(-(u-x)^2/s^2) du = \sum_{j=0}^{n} b_j(\Gamma_j) \cdot x^j \cdot \frac{1}{(\sigma^2+\varepsilon \cdot s^2)^{1/2}} \cdot \frac{1}{\sqrt{\pi}} \cdot \exp(-x^2/(\sigma^2+\varepsilon \cdot s^2)) \\ \Gamma_j &= \Gamma(\tfrac{j+1}{2})\end{aligned}\right\}. \quad (31)$$

For odd j the coefficients $b_j$ are zero, if the power n is even, whereas for odd n all coefficients with even j vanish.

We now regard the set of test functions g(x) given by a power expansion in x multiplied with the above Gaussian, i.e.:

$$g(x) = \sum_{j=0}^{n} a_j \cdot x^j \cdot \frac{1}{\sigma \cdot \sqrt{\pi}} \cdot \exp(-x^2/\sigma^2). \quad (32)$$

With the help of the forgoing results the general solution of the related convolution problem the following expansion can be stated:

$$\hat{g}(x) = \sum_{j=0}^{n} c_j \cdot x^j \cdot \frac{1}{\sqrt{\pi}} \cdot \frac{1}{(\sigma^2+\varepsilon \cdot s^2)^{1/2}} \cdot \exp(-x^2/(\sigma^2 + \varepsilon \cdot s^2)). \quad (33)$$

This resulting set of functions is important with regard to the case n → ∞ in $L_2(-\infty, \infty)$, since $c_n$ is proportional to Γ(n/1+1/2). Therefore the additional proportionality of $c_n$ (n → ∞) to $a_n$ → 0 (n → ∞) has to be satisfied to save the convergence of ĝ and the existence of the norm requested by $L_2$.

Since a Gaussian test function g(x) is rapidly decreasing, we anticipate that the problem of the function g(x) has to incorporate the opposite behavior. We possess some different toolkits to manage this task. Thus we can use equations (4) and (13 – 13d) for the determination of G from the test function g, which, at first, we restrict to a single Gaussian again:

$$G = O^{-\varepsilon} \cdot g; \quad \varepsilon = -1. \tag{34}$$

$$\rightarrow G = \int_{-\infty}^{\infty} K^{-1}(s, u-x) \cdot g(u) du = O^{-2\varepsilon} \int_{-\infty}^{\infty} K(s, u-x) \cdot g(u) du. \tag{34a}$$

The question is now, which possible procedure to determine is the least intricate one. Equation (34) solves the problem by repeated differentiations as already introduced. The determination of G via equation (4) at the left-hand side of equation (34a) requires a lot of effort. On the right-hand side of equation (34a) we perform at first the convolution with K and thereafter the operation with $O^2$. This procedure leads to a quite interesting access to the inverse problem: In a rather different connection Feynman et al [32] used with regard to integrations in path integral quantization the formula:

$$K(\varepsilon, s, u-x) = \frac{1}{\sqrt{\pi \cdot \varepsilon}} \cdot \frac{1}{s} \cdot \exp(-(u-x)^2/(\varepsilon \cdot s^2)). \tag{35}$$

If $\varepsilon = -1$ and $g$ given according above we obtain the desired kernel for $K^{-1}$. Thus the final result is:

$$\left. G = \frac{1}{\sqrt{\pi}} \cdot \frac{1}{\sqrt{\sigma^2 + \varepsilon \cdot s^2}} \cdot \exp(-x^2/(\sigma^2 + \varepsilon \cdot s^2)) \right|_{\varepsilon = -1}. \tag{36}$$

Thus convergence is only obtained, if $\sigma^2 > s^2$. If $\sigma = s$ the solution is the δ-function, and for $\sigma < s$ the solution is complex-valued. With regard to Gaussian convolution/deconvolution we can formally write:

$$O^{-\varepsilon} \cdot g(x) = \frac{1}{\sqrt{\pi}} \cdot \frac{1}{\sqrt{\sigma^2 + \varepsilon \cdot s^2}} \cdot \exp(-x^2/(\sigma^2 - \varepsilon \cdot s^2)) = \frac{1}{\sqrt{\pi}} \cdot \frac{1}{\sqrt{\sigma^2 + \varepsilon \cdot s^2}} \cdot \exp(-\sigma^2 \cdot x^2 \cdot \sum_{n=0}^{\infty} (-\varepsilon)^n \cdot (s^{2n}/\sigma^{2n})). \tag{37}$$

The inverse problem of the function g(x) according to equation (32), i.e. a Gaussian multiplied with the power $x^n$, follows the same principle, and we have only to perform the substitution $\varepsilon = -1$ in equation (37). It is obvious that a Gaussian convolution also converges, if $s > \sigma$; the resulting *rms*-value is always $s' = (s^2 + \sigma^2)^{1/2}$. The back calculation also exists, since we have account for $s' > s$ in order to obtain the initial Gaussian given by the *rms*-value σ.

A rather peculiar behavior shows the function;

$$f(x) = \exp(-|\mu \cdot x|). \tag{38}$$

Thus both possible ways to calculate the Gauss-transformed by either $O^{-1}$ or convolution with K provide the solution:

$$\hat{f} = \exp(-|\mu \cdot x|) \cdot \exp(\tfrac{1}{4} \cdot s^2 \cdot \mu^2). \tag{39}$$

It should be pointed out that the operator $O^{-1}$ as well as the operator $O$ only contains derivations of even order and continuity is always guaranteed at $x = 0$. Only differential operators of odd order lead to a jump at $x = 0$. The deconvolution of a given $f(x)$ can be performed with all calculation procedures presented here in a rather simple manner; the result is:

$$\bar{f}(x) = O \cdot f(x) = O^2 \cdot \hat{f} = \int_{-\infty}^{\infty} K^{-1}(s, u-x) \cdot f(u) du = \exp(-|\mu \cdot x|) \cdot \exp(-\tfrac{1}{4} \cdot s^2 \cdot \mu^2). \tag{40}$$

The resulting solution of deconvolution according to equation (40) can be subjected to a Laplace transform:

$$\int_{0}^{\infty} \exp(-|\mu \cdot x|) \cdot \exp(-\tfrac{1}{4} \cdot s^2 \cdot \mu^2) d\mu = \tfrac{\sqrt{\pi}}{s} \exp(x^2 / s^2) \cdot erfc(\tfrac{x}{s}). \tag{41}$$

A Laplace transform of the image function of equation (39) does not exist, which is induced by the term $\exp(0.25 \cdot s^2 \cdot \mu^2)$. An extension of equation (39) by polynomials can be solved with the help of the already discussed methods:

$$f(x) = \exp(-|\mu \cdot x|) \cdot x^n. \tag{42}$$

The substitution (27) can be applied, too. The resulting solutions are of the types (39) or (40), but, in addition, with a sequence of powers with regard to $x$:

$$\bar{f}(x) = \exp(-|\mu \cdot x|) \cdot \exp(\varepsilon \cdot \tfrac{1}{4} \cdot s^2 \cdot \mu^2) \cdot P_n(\varepsilon, x); \quad (\varepsilon = \pm 1). \tag{42a}$$

In a previous paper [7] we have also stated a further possible representation of the inverse kernel $K^{-1}$ with some advantages in numerical calculations of inverse problems of one Gaussian kernel:

$$K^{-1}(s, u-x) = \delta(u-x) + (O^2 - O) \cdot K(s, u-x) = \delta(u-x) + \sum_{n=1}^{\infty} (-1)^n \cdot s^{2n} \cdot \frac{2^n - 1}{n! \cdot 4^n} \cdot (d^{2n}/dx^{2n}) \cdot K(s, u-x). \tag{43}$$

This formula results from the identities $O^2 \cdot K(s, u-x) = K^{-1}(s, u-x) =$ and $O \cdot K(s, u-x) = \delta(u-x)$. By taking account for the power expansion of $O^2$ and $O$ and the corresponding subtraction $O^2 - O$ in the above equation (43) the unit operator '1' related to K is cancelled and a modified calculation procedure is obtained. Due to the δ-distribution the lowest term of deconvolution is the identity of source and image function. The integral notation of this equation assumes the shape:

$$\rho = \varphi + \sum_{n=1}^{\infty} (-1)^n \cdot s^{2n} \cdot \frac{2^n - 1}{n! \cdot 4^n} \cdot (d^{2n}/dx^{2n}) \cdot \int \varphi(u) \cdot K(s, u - x) du. \tag{44}$$

If the image function φ belongs to $C^\infty$, the use of Hermite polynomials is not requested.

## 2.2. Inverse problem according to IFIE2 and LNS method

As already pointed out the main purpose here is the inverse problem of linear combinations of Gaussian kernels. In order to reduce the effort of formula writing, we restrict ourselves to maximal three combinations. The operator formulation analog to equations (5 - 6) of this convolution reads (in one dimension):

$$\varphi(x) = O_g^{-1} \cdot \rho(x) = [c_0 \cdot O_0^{-1}(s_0) + c_1 \cdot O_1^{-1}(s_1) + c_2 \cdot O_2^{-1}(s_2)] \cdot \rho(x). \tag{45}$$

It is the task to determine ρ(x), if φ(x) is given, which is equivalent to determine $K_g^{-1}$. In every case, the condition $O_g^{-1} \cdot O_g = 1$ has to be satisfied.

In a previous study [5] we have made use of the operator calculus to determine $O_g$ and via $O_g^{-1}$ to derive the inverse kernel $K_g^{-1}$. The operator calculus provides the following relationship:

$$O_g \cdot O_g^{-1} = 1 = O_g \cdot [c_0 \cdot O_0^{-1} + c_1 \cdot O_1^{-1} + c_2 \cdot O_2^{-1}] \Rightarrow O_g = [c_0 \cdot O_0^{-1} + c_1 \cdot O_1^{-1} + c_2 \cdot O_2^{-1}]^{-1} \tag{46}$$

We have now to evaluate the following Lie series of the operator function $O_g$ in terms of the operators $O_0^{-1}$ and $O_1$:

If the image function φ belongs to $C^\infty$, the use of Hermite polynomials is not requested.

$$O_g = [c_0 \cdot O_0^{-1} + c_1 \cdot O_1^{-1} + c_2 \cdot O_2^{-1}]^{-1}. \tag{47}$$

We use the following relation for commutative operators [32]:

$$[A + B]^{-1} = \sum_{n=0}^{\infty} (-1)^n \cdot A^{-n-1} \cdot B^n. \tag{48}$$

With the help of the substitutions A = $c_0 \cdot O_0^{-1}$ and B = $c_1 \cdot O_1^{-1} + c_2 \cdot O_2^{-1}$ we are able to derive the operator function $O_g$, which satisfies $O_g \cdot \varphi = \rho$, and the related inverse kernel $K_g^{-1}$:

$$\varphi = [c_0 \cdot O_0^{-1} + c_1 \cdot O_1^{-1} + c_2 \cdot O_2^{-1}] \cdot \rho. \tag{49}$$

$$\Rightarrow \frac{1}{c_0} \cdot O_0 \cdot [1 + \frac{c_1}{c_0} \cdot O_0 \cdot O_1^{-1} + \frac{c_2}{c_0} \cdot O_0 \cdot O_2^{-1}]^{-1} \cdot \varphi = \rho. \tag{50}$$

In view of the following section, we introduce the abbreviation:

$$f(x,y,z) = \frac{1}{c_0} \cdot \int K_0^{-1}(s_0, \vec{u} - \vec{x}) \cdot \varphi(\vec{u}) d^3 u. \tag{51}$$

Function $f$ incorporates the inhomogeneous part of the Fredholm integral equation of second kind (IFIE2). Since this method differs from the previous publication [5], this section is dedicated to this task.

In order to derive an alternative method to solve the inverse problem of a linear combination of Gaussian convolutions, we consider equation (50) with regard to two kernels (the generalization to $c_2 \neq 0$ will be stated thereafter), we multiply with $O_0/c_0$ from the left-hand side. By that, we readily obtain the desired formula, which will be transformed to an IFIE2:

$$\frac{1}{c_0} \cdot O_0 \cdot \varphi = [1 + \frac{c_1}{c_0} \cdot O_0 \cdot O_1^{-1}] \cdot \rho. \tag{52}$$

We should like to point out that the preceding equations (46 – 52) result from a power expansion of the expression $[1 + (c_0/c_1) \cdot O_0 \cdot O_1^{-1}]^{-1}$ in terms of a Lee series in order to resolve equation (52) with regard to $\rho$. However, equation (52) can immediately be transformed to an integral equation by the principles elaborated above, i.e. the left-hand side implies a deconvolution term of the operator $O_0$ applied to $\varphi$, whereas the operator $O_0 \cdot O_1^{-1}$ implies a convolution term $K_f$ ($s_1 > s_0$) applied to $\rho$:

$$\left. \begin{array}{l} f(\vec{x}) = \rho(\vec{x}) + \frac{c_1}{c_0} \cdot \int \rho(\vec{u}) \cdot K_f(\sigma, \vec{u} - \vec{x}) d^3 u \\ \sigma^2 = s_1^2 - s_0^2 \end{array} \right\}. \tag{53}$$

With the substitution $\lambda = -(c_1/c_0)$ equation (53) represents the usual notation of an IFIE2; the inhomogeneous term $f$ results from a deconvolution procedure and $K_f(\sigma, u - x)$ is a normalized Gaussian kernel with regard to the parameter $\sigma$ in equation (53). The inverse problem is solved by finding the solution of equation (53), which can be done best with the help of LNS, i.e. the iterated kernel $K_{f(n)}$ has to be determined from the above kernel $K_f(\sigma, u-x)$. The $n^{th}$ – iterated kernel is calculated by the procedure:

$$K_{f_{(n)}}(\vec{u} - \vec{x}) = \int\int \ldots \int K_f(\sigma, \vec{u} - \vec{u}_1) \cdot K_f(\sigma, \vec{u}_1 - \vec{u}_2) \cdot \ldots \cdot K_f(\sigma, \vec{u}_{n-1} - \vec{x}) d^3 u_1 d^3 u_2 \ldots d^3 u_{n-1}. \tag{54}$$

The resolving kernel $K_{res}$ is given by:

$$\left. \begin{array}{l} K_{res}(\vec{u} - \vec{x}, \lambda) = \sum_{n=0}^{L} \lambda^n \cdot K_{f_{(n+1)}}(\vec{u} - \vec{x}) \\ L \to \infty \end{array} \right\}. \tag{55}$$

The solution of the integral equation becomes:

$$\rho(\vec{x}) = \int K_{res}(\vec{u}-\vec{x},\lambda) \cdot f(\vec{u}) d^3u \ . \tag{56}$$

In practical applications, we have to be aware of a *finite limit L* in equation (55), and L → ∞ cannot be carried out. The evaluation of the iterated terms $K_{f(n)}$ is rather simple, since $K_f$ is the normalized Gaussian kernel. Thus $K_{f(1)}$ is the normalized Gaussian kernel itself. $K_{f(2)}$ results from a composite convolution:

$$K_{f(h)} = \int K_f(\sigma,\vec{u}_1-\vec{x}) \cdot K_f(\tau,\vec{u}_1-\vec{u}) d^3u_1 = \frac{1}{\sqrt{\pi^3}} \cdot \frac{1}{\sqrt{(\sigma^2+\tau^2)^3}} \cdot \exp(-(\vec{u}-\vec{x})^2/(\sigma^2+\tau^2)). \tag{57}$$

$$\text{If } \sigma^2 = \tau^2: \quad K_{f(2)} = \frac{1}{\sqrt{\pi^3}} \cdot \frac{1}{(\sqrt{2\sigma^2})^3} \cdot \exp(-(\vec{u}-\vec{x})^2/2\sigma^2). \tag{58}$$

In equation (57) we have introduced the 'helping formula' $K_{f(h)}$, which allows us to determine $K_{f(3)}$, $K_{f(4)}$…, by applying equation (57) iteratively. Thus by the fixation $\tau^2 = 2\sigma^2$ we obtain via equation (57) $K_{f(3)} = K_f(3\sigma^2)$. In the same fashion $K_{f(4)}$ is determined by $K_f(4\sigma^2)$ and $K_{f(n)}$ by $K_f(n\sigma^2)$. $K_{f(n)}$ appears in every order of the calculation procedure with the help *LNS*.

As already mentioned, rapid convergence is reached, if $c_0 \gg c_1$ and the ratio $\lambda$ is small. Then the powers of $\lambda$ become correspondingly much smaller. Thus for $c_0 = 0.9$ and $c_1 = 0.1$ we obtain $\lambda = -0.11111$ ($\lambda^2 = 0.01234$), whereas for $c_0 = 0.55$ and $c_1 = 0.45$ we obtain $\lambda = -0.8181$ and $\lambda^2 = 0.66942$. There is also a principal difference between the two calculation procedures with regard to the parameters $s_0$ and $s_1$. The application of the LNS method only requires $\sigma^2 > 0$, i.e. $s_1^2 > s_0^2$, while the previous method [5] only exists, if $s_1^2 > 2s_0^2$. A further difference between the two methods refers to the inverse kernel $K_g^{-1}$, which has to be determined in the first method to calculate the source function ρ from a given image function φ, whereas via LNS method we can directly calculate the source function $\rho$ from a given image function $\varphi$ without determination of the inverse kernel. The extension to a linear combination of three Gaussian convolution kernels leads with regard to the inverse problem to the following IFIE2:

$$\left.\begin{array}{l} f(\vec{x}) = \rho(\vec{x}) + \frac{c_1}{c_0} \cdot \int \rho(\vec{u}) \cdot K(\sigma_1,\vec{u}-\vec{x}) d^3u + \frac{c_2}{c_0} \cdot \int \rho(\vec{u}) \cdot K(\sigma_2,\vec{u}-\vec{x}) d^3u \\ \sigma_1^2 = s_1^2 - s_0^2; \quad \sigma_2^2 = s_2^2 - s_0^2 \end{array}\right\} . \tag{59}$$

In order to evaluate equation (59) by equation (54), we write this equation in the form:

$$\left.\begin{array}{l} f(x) = \rho(x) - \lambda \cdot \int \rho(u) \cdot [K(\sigma_1,u-x) + \alpha \cdot K(\sigma_2,u-x)] du \\ \lambda = -c_1/c_0; \quad \alpha = c_2/(c_0 \cdot c_1) \end{array}\right\} . \tag{60}$$

For the evaluation of the inverse kernel we need to calculate $K_{f(n)}$:

$$K_{f_{(n)}} = \int\int\cdots\int \left\{ \begin{array}{l} [K_f(\sigma_1, \vec{u}-\vec{u}_1) + \alpha \cdot K_f(\sigma_2, \vec{u}-\vec{u}_1)] \cdot [K_f(\sigma_1, \vec{u}_1-\vec{u}_2) + \alpha \cdot K_f(\sigma_2, \vec{u}_1-\vec{u}_2)] \cdot \\ \cdots \cdot [K_f(\sigma_1, \vec{u}_{n-1}-\vec{x}) + \alpha \cdot K_f(\sigma_2, \vec{u}_{n-1}-\vec{x})] d^3u_1 d^3u_2 \cdots d^3u_{n-1} \end{array} \right\}. \quad (61)$$

It is evident that $K_{f(n)}$ has to contain the terms $K_f(n \cdot \sigma_1, u-x)$ and $\alpha^n \cdot K_f(n \cdot \sigma_2, u-x)$, but the binominal theorem also provides mixed products, and by evaluation of equation (61) $K_{f(n)}$ assumes the shape:

$$K_{f_{(n)}}(\vec{u}-\vec{x}) = \sum_{j=0}^{n} \alpha^j \cdot \binom{n}{j} \cdot K_f((n-j) \cdot \sigma_1^2 + j \cdot \sigma_2^2, \vec{u}-\vec{x}). \quad (62)$$

It must be pointed out that now equation (55) has to be evaluated with the help of equation (61). With regard to convergence aspects in the above cases according to equations (57 – 61) it is obvious that convergence is fast, if $c_0$ satisfies $c_0 \gg c_1$ or $c_0 \gg c_2$, i.e. the leading term refers to $c_0$ and the additional contributions only represent (small) long-range tails. Please note that the LNS method is also applicable, if $c_1 < 0$ ($c_0 + c_1 + c_2 = 1$) is assumed. Examples for this case will be presented in the section 3.

## 2.3. Theoretical aspects of Bethe-Bloch equation (BBE)

The application of the Bethe-Bloch equation (BBE) for the determination of the electronic stopping power is established for the passage of electrons and protons through homogeneous media. A particular importance of BBE appears in Monte-Carlo calculations to simulate behavior of charged projectile particles along the track. This equation reads:

$$\left. \begin{array}{l} -dE(z)/dz = (K/v^2) \cdot [\ln(2mv^2/E_I) - \ln(1-\beta^2) + \\ \quad + a_{shell} + a_{Barkas} + a_0 v^2 + a_{Bloch}] \\ K = (Z\rho/A_N) \cdot 8\pi q^2 e_0^4 / 2m \end{array} \right\}. \quad (63)$$

$E_I$ is the atomic ionization energy, weighted over all possible transition probabilities of atomic/molecular shells, $\beta = v/c$, q denotes the charge number of the projectile (e.g. proton, carbon ion), and Z, $\rho$ and $A_N$ refer to the charge, density and relative mass number of the absorbing medium. According to ICRU49 [41] we have to put $a_0 = -1$. The meaning of the correction terms $a_{shell}$, $a_{Barkas}$, $a_0$ and $a_{Bloch}$ is explained in [38 – 43]. In this study we mainly consider the basic aspects of the Barkas effect. A theory of this effect has been developed in [44].

The Barkas effect represents a correction of BBE due to the electron capture of the positively charged protons at lower energies in the domain of the Bragg peak and behind leading to a slightly increased range $R_{csda}$, whereas the negatively charged anti-protons cannot capture electrons from the environmental electrons. Therefore their range is slightly smaller. With regard to protons this kind of correction works, i.e. the charge $q^2 = 1$ is assumed along the total proton track, whereas for charged ions such as He or $C^6$ is appears to be insufficient to keep the nuclear charge constant along the total track and to restrict the electron capture only to the small Barkas correction [45]. This means that all positively charged projectile particles stand in permanent exchange of energy E and charge q with environment, and, as a consequence, $q^2$ is a function of the actual residual energy, i.e. $q^2 = q^2(E)$, and only for $E = E_0$ (initial energy) $q^2 = q_0^2$ is valid. A correct modification of BBE by accounting for $q^2(E)$ due to electron capture makes the Barkas correction superfluous.

A further critical aspect of BBE, which leads to a modification by accounting for $q^2(E)$ is the range $R_{csda}$ of the electronic stopping power. Thus a naïve application of BBE would lead to the conclusion that a carbon ion would require the initial energy per nucleon $E_0$ (carbon ion) = 3 x $E_0$(proton), since the square of the carbon charge amounts to 36 and the nuclear mass unit is 12 x nuclear mass unit of the proton. However, the ratio is not 3 to obtain the same range $R_{csda}$, but about 25/12. The Monte-Carlo code GEANT4 assumes an average charge $q_{Average} = 5.06$ for the simulations of the carbon tracks. This is, however, not satisfactory, since electron capture is a dynamical process. Therefore the range of charged particles has been subjected to many studies due to the increasing importance of carbon ions in radiotherapy [41 – 51]. It is also possible to substitute the electron mass m by the reduced mass m $\Rightarrow$ µ. However, this leads for protons to a rather small correction (i.e., less than 0.1 % for protons). For complex systems $E_I$ and some other contributions like $a_{shell}$ and $a_{Barkas}$ can only be approximately calculated by simple quantum-mechanical models (e.g., harmonic oscillator); the latter terms are often omitted and $E_I$ is treated as a fitting parameter, but different values are proposed and used [41]. The restriction to the logarithmic term leads to severe problems, if either v → 0 or $2m v^2 /E_I$ → 1. It should be added that a correct treatment of the electron capture removes the singularity of positively charged ions, since $q^2(E)$ → 0, if the residual energy E assumes zero.

In previous publications [6, 8, 62] we have presented an analytical integration of BBE, which is the physical base in the transport of protons and other charged particles such as heavy carbon ions.

In order to obtain the integration of BBE, we start with the logarithmic term and perform the substitutions:

$$v^2 = 2E/M; \quad \beta_I = 4m/ME_I; \quad E = (1/\beta_I)\exp(-u/2) \,. \quad (64)$$

With the help of substitution (and without any correction terms), BBE leads to the integration:

$$-\int du \exp(-u) \cdot (1/u) = \tfrac{1}{2} K \cdot \beta_I^2 \cdot M \int dz$$
$$K = (Z\rho / A_N) \cdot 8\pi q^2 e_0^4 / 2m \qquad (65)$$

The boundary conditions of the integral are:

$$z = 0 \Rightarrow E = E_0 \ (\text{or}: u = -2\ln(E_0 \cdot \beta_I))$$
$$z = R_{CSDA} \Rightarrow E = 0 \ (\text{or}: u \Rightarrow \infty) \qquad (66)$$

The general solution is given by the Euler exponential integral function Ei($\xi$) with P.V. = principal value:

$$\tfrac{1}{2} K \cdot M \cdot \beta_I^2 \cdot R_{CSDA} = -P.V. \int_{-\xi}^{\infty} u^{-1} \exp(-u) du = Ei(\xi)$$
$$\xi = 2\ln(4mE_0 / ME_I) \quad \text{and} \quad \xi > 0 \qquad (67)$$

Some details of Ei($\xi$) and its power expansions can be found in [35]. The critical case $\xi = 0$ results from $E_{critical} = ME_I/4m$ (for water with $E_I = 75.1$ eV, the critical energy $E_{critical}$ amounts to 34.474 keV; for Pb with $E_I \approx 800$ eV to about 0.4 MeV). Since the logarithmic term derived by Bethe implies the Born approximation, valid only if the transferred energy $E_{transfer} \gg$ the energy of shell transitions, the above corrections, exempting the Bloch correction, play a significant role in the environment of the Bragg peak, and the terms $a_0$, $a_{shell}$, and $a_{Barkas}$ remove the singularity. However, the integration procedure according to the above equation (67) remains valid, if we account for the correction terms. With respect to numerical integrations (Monte Carlo), we note that, in the environment of $E = E_{critical}$, the logarithmic term may become crucial (leading to overflows); rigorous cutoffs circumvent the problem. Therefore, the shell corrections are an important feature for low proton energies.

The result of the integration yields a power expansion for $R_{CSDA}$ in terms of $E_0$:

$$R_{CSDA} = \frac{1}{\rho} \cdot \frac{A_N}{Z} \sum_{n=1}^{N} \alpha_n E_I^{pn} E_0^n \quad (N \Rightarrow \infty). \qquad (68)$$

The coefficients $\alpha_n$ are determined by the integration procedure and only depend on the parameters of the BBE. For applications to therapeutic protons, i.e., $E_0 < 300$ MeV, a restriction to $N = 4$ provides excellent results (Figure 1). For water, we have to take $E_I = 75.1$ eV, $Z/A_N = 10/18$, $\rho = 1$ g/cm$^3$; formula (68) becomes:

$$R_{CSDA} = \sum_{n=1}^{N} a_n E_0^{\,n} \quad (N \Rightarrow \infty) \qquad (69)$$

The values of the parameters of Formulas with restriction to $N = 4$ are displayed in Tables 1 and 2.

**Table 1.** Parameter values for equation (68) if $E_0$ is in MeV, $E_I$ in eV and $R_{CSDA}$ in cm.

| $\alpha_1$ | $\alpha_2$ | $\alpha_3$ | $\alpha_4$ | p1 | p2 | p3 | p4 |
|---|---|---|---|---|---|---|---|
| $6.8469 \cdot 10^{-4}$ | $2.26769 \cdot 10^{-4}$ | $-2.4610 \cdot 10^{-7}$ | $1.4275 \cdot 10^{-10}$ | 0.4002 | 0.1594 | 0.2326 | 0.3264 |

**Table 2.** Parameter values for equation (69), if $E_0$ is in MeV, $E_I$ in eV and $R_{CSDA}$ in cm.

| $a_1$ | $a_2$ | $a_3$ | $a_4$ |
|---|---|---|---|
| $6.94656 \cdot 10^{-3}$ | $8.13116 \cdot 10^{-4}$ | $-1.21068 \cdot 10^{-6}$ | $1.053 \cdot 10^{-9}$ |

The determination of $A_N$ and $Z$ is not a problem in case of atoms or molecules, where weight factors can be introduced according to the Bragg rule; for tissue heterogeneities, it is already a difficult task. Much more difficult is the accurate determination of $E_I$, which results from transition probabilities of all atomic/molecular states to the continuum ($\delta$-electrons). Thus with regard to stopping powers of protons in different media according to [41], there are sometimes different values of $E_I$ proposed (e.g., for Pb: $E_I = 820$ eV and $E_I = 779$ eV). If we use the average (i.e., $E_I = 800.5$ eV), the above formula provides a mean standard deviation of 0.27 % referred to stopping-power data in [41], whereas for $E_I = 820$ eV or $E_I = 779$ eV we obtain 0.35 % - 0.4 %. If we apply the above formula to data of other elements listed in [41], the mean standard deviations also amount to about 0.2 % - 0.4 %.

Instead of the usual power expansion (69), we can represent all integrals in terms of Gompertz-type functions multiplied with a single exponential function by collection of all exponential functions obtained by equations (63 – 67) and the substitution $\beta_l E = \exp(-u/2)$. The Gompertz-function is defined by:

$$\left. \begin{aligned} \exp(-\xi \exp(-u/2)) &= 1 - \xi \exp(-u/2) + \tfrac{1}{2!}\xi^2 \exp(-u). - .. + = \\ &= 1 + \sum_{k=1}^{\infty} \tfrac{1}{k!}(-1)^k \xi^k \exp(-ku/2) \end{aligned} \right\}. \quad (70)$$

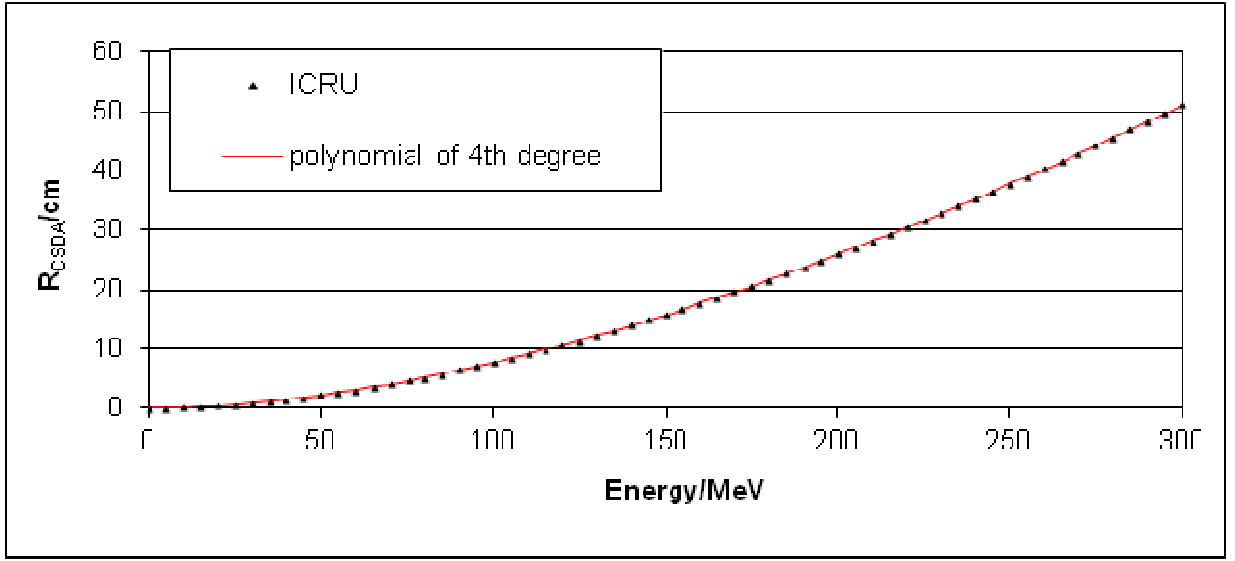

**Figure 1.** Comparison data in [41] of proton $R_{CSDA}$ range (up to 300 MeV) in water and the fourth-degree polynomial (equation 69). The average deviation amounts to 0.0013 MeV.

By inserting the integration boundaries $u = 2 \cdot \ln \cdot 4m \cdot E_0/(M \cdot E_I)$, i.e., $E = E_0$ and $u \to \infty$ ($E = 0$), the integration leads to a sequence of exponential functions; the power expansion is replaced by:

$$R_{CSDA} = a_1 E_0 \cdot [1 + \sum_{k=1}^{N}(b_k - b_k \exp(-g_k \cdot E_0))] \quad (N \Rightarrow \infty). \quad (70a)$$

For therapeutic protons, the restriction to $N = 2$ provides the same accuracy (Figure 2) as formula (69); the parameters are given in Table 3 ($a_1$ is the same as in Table 2).

**Table 3.** Parameters of Formula (70a); $b_1$ and $b_2$ are dimensionless; $g_1$ and $g_2$ are given in MeV$^{-1}$.

| $b_1$ | $b_2$ | $g_1$ | $g_2$ |
|---|---|---|---|
| 15.14450027 | 29.84400076 | 0.001260021 | 0.003260031 |

In the following, we shall verify that the latter formula provides some advantages with respect to the inversion $E_0 = E_0(R_{CSDA})$.

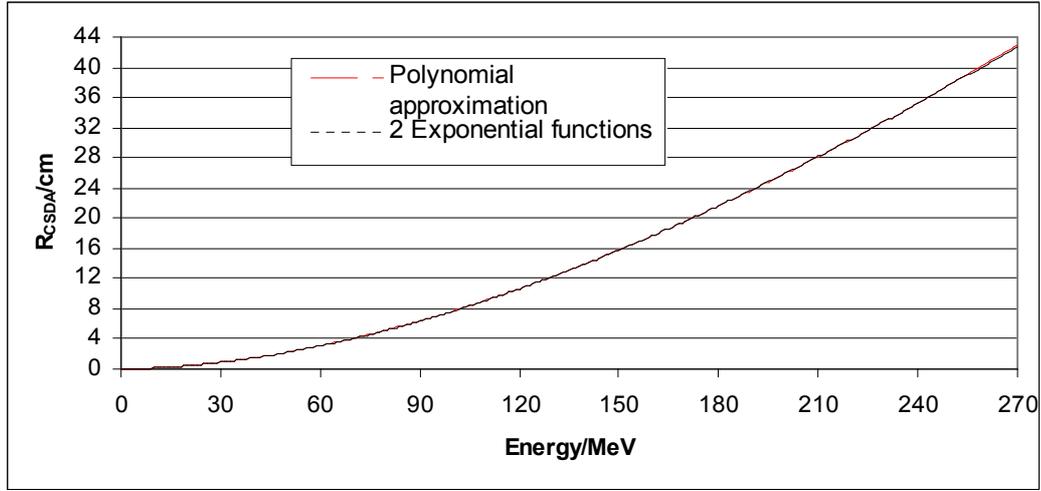

**Figure 2.** $R_{CSDA}$ calculation - comparison between a fourth-degree polynomial (equation (69)) and two exponential functions (equation (70a)).

Above formulas can also be used for the calculation of the residual distance $R_{CSDA} - z$, relating to the residual energy $E(z)$; we have only to perform the substitutions $R_{CSDA} \rightarrow R_{CSDA} - z$ and $E_0 \rightarrow E(z)$ in these formulas. In various problems, the determination of $E_0$ or $E(z)$ as a function of $R_{CSDA}$ or $R_{CSDA} - z$ is an essential task. The power expansion implies again a corresponding series $E_0 = E_0(R_{CSDA})$ in terms of powers:

$$\left. \begin{array}{l} E_0 = \sum_{k=1}^{\infty} c_k R_{CSDA}^{\ k} \\ c_1 = 1/a_1, \ \ c_2 = -a_2/a_1^3, \ \ c_3 = (2a_2^2 a_1^{-1} - a_3)/a_1^4, \dots \\ c_k = f(a_k, a_{k-1}, a_{k-2}\dots)/a_1^{k+1} \ \ (k > 3) \end{array} \right\}. \quad (71)$$

The coefficients $c_k$ are calculated by a recursive procedure; we have given the first three terms in formula (71). Due to the small value of $a_1 = 6.8469 \cdot 10^{-4}$, this series is ill-posed, since there is no possibility to break off the expansion; it is divergent and the signs of the coefficients $c_k$ are alternating, see [35]. The inversion procedure of this equation leads to the formula (see e.g.[6]):

$$\left. \begin{array}{l} E_0 = R_{csda} \sum_{i=1}^{N} c_k \exp(-\lambda_k R_{csda}) \ \ (N \rightarrow \infty) \\ E(z) = (R_{csda} - z) \sum_{i=1}^{N} c_k \exp(-\lambda_k (R_{csda} - z)) \end{array} \right\}. \quad (72)$$

By extending formula (72) to different media the inverse formula of equation (70a) becomes:

$$\left. \begin{array}{l} c'_k = c_k \cdot (18/10) \cdot Z \cdot \rho \cdot (75.1/E_I)^{q_k} / (A_N \cdot \rho_w) \\ \lambda^{-1}{}'_k = \lambda^{-1}{}_k \cdot (10 \cdot \rho_w / 18) \cdot (75.1/E_I)^{p_k} \cdot A_N / (\rho \cdot Z) \\ E(z) = (R_{CSDA} - z) \cdot \sum_{k=1}^{5} c'_k \cdot \exp[-(R_{CSDA} - z) \cdot \lambda'_k] \end{array} \right\}. (73)$$

For therapeutic protons, a very high precision is obtained by the restriction to N = 5 (Table 4 and Figure 4). Formula (72) is also suggested by a plot $S(R_{CSDA}) = E_0(R_{CSDA})/R_{CSDA}$ according to equation (72). This plot is shown in Figure 3 and gives rise for an expansion of $S(R_{CSDA})$ in terms of exponential functions. This plot is obtained by an interchange of the plot $E_0$ versus $R_{CSDA}$ and a calculation according to the above relation.

**Table 4.** Parameters of the inversion formula (73) with N = 5 (dimension of $c_k$: cm/MeV, $\lambda_k$: cm$^{-1}$).

| $c_1$ | $c_2$ | $c_3$ | $c_4$ | $c_5$ | $\lambda_1^{-1}$ | $\lambda_2^{-1}$ | $\lambda_3^{-1}$ | $\lambda_4^{-1}$ | $\lambda_5^{-1}$ |
|---|---|---|---|---|---|---|---|---|---|
| 96.63872 | 25.0472 | 8.80745 | 4.19001 | 9.2732 | 0.0975 | 1.24999 | 5.7001 | 10.6501 | 106.72784 |

| $P_1$ | $P_2$ | $P_3$ | $P_4$ | $P_5$ | $q_1$ | $q_2$ | $q_3$ | $q_4$ | $q_5$ |
|---|---|---|---|---|---|---|---|---|---|
| -0.1619 | -0.0482 | -0.0778 | 0.0847 | -0.0221 | 0.4525 | 0.195 | 0.2125 | 0.06 | 0.0892 |

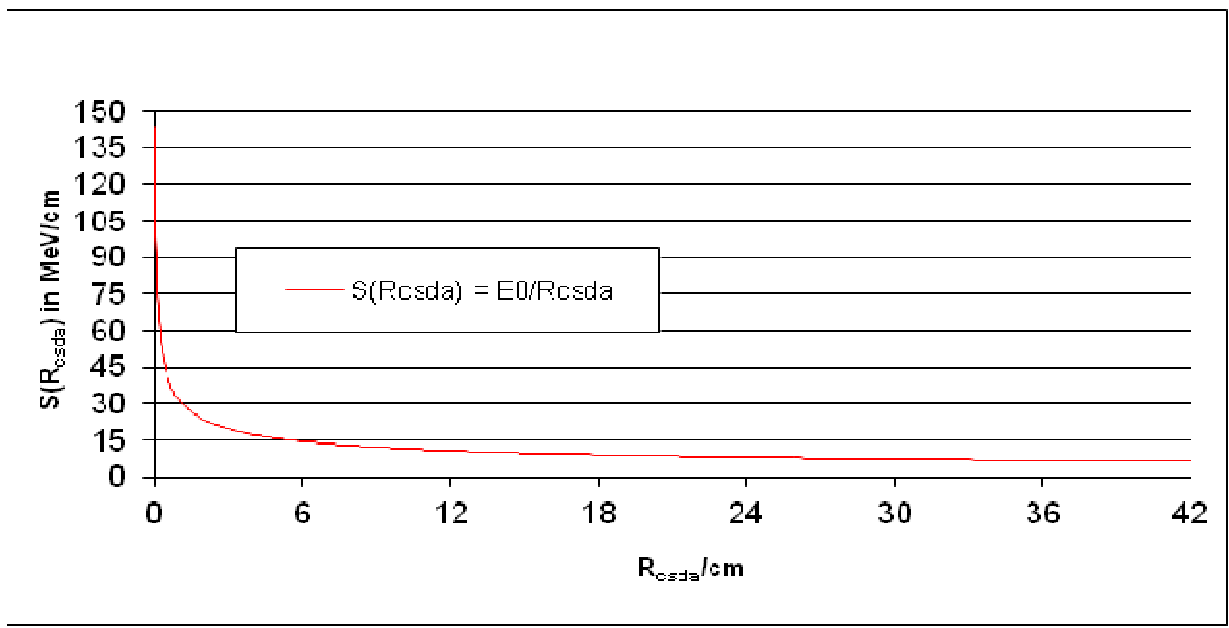

**Figure 3.** Plot $S(R_{CSDA}) = E_0/R_{CSDA}$ provides a justification of the representation of S by exponential functions.

One way to obtain the inversion Formula is to find $S(R_{CSDA})$ by a sum of exponential functions with the help of a fitting procedure. Thus it turned out that the restriction to five exponential functions is absolutely sufficient and yields a very high accuracy. A more rigorous way (mathematically) has been described in the LR of [62].

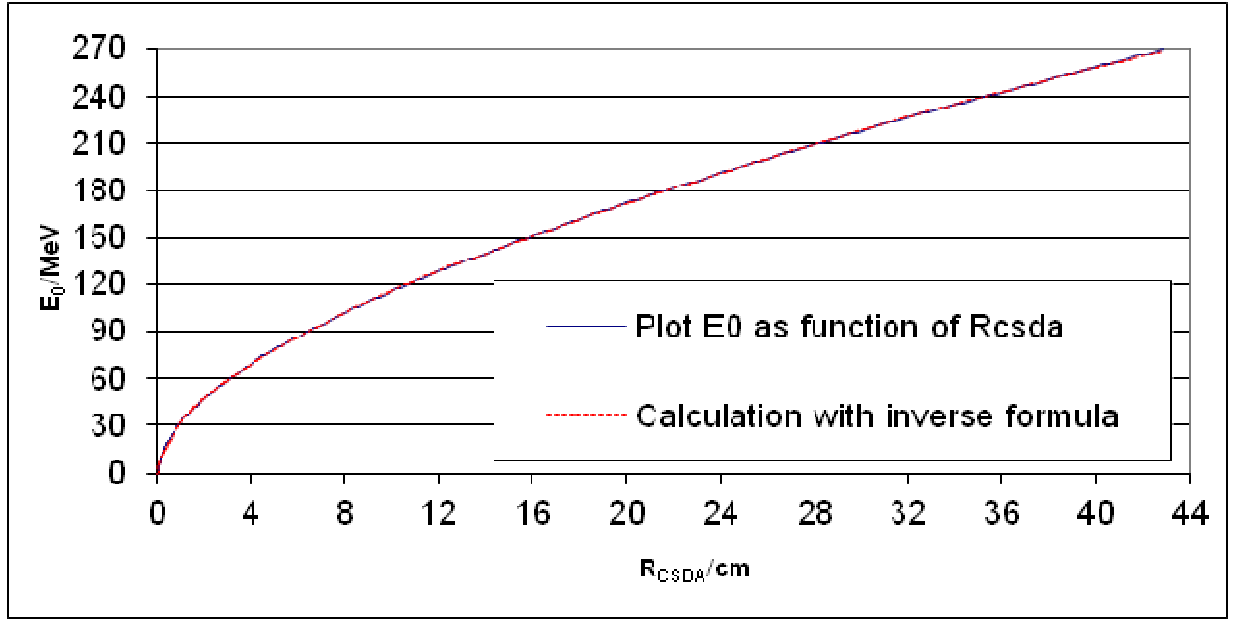

**Figure 4.** Test of the inverse Formula (40) $E_0 = E_0(R_{CSDA})$ by five exponential functions. The mean deviation amounts to 0.11 MeV. The plot results from Figure 1.

The residual energy $E(z)$, appearing in equation (73), is the desired analytical base for all calculations of stopping power and comparisons with GEANT4. The stopping power is determined by $dE(z)/dz$ and yields the following expression:

$$\left.\begin{aligned}
S(z) &= dE(z)/dz \\
&= -E(z)/(R_{CSDA} - z) + \sum_{k=1}^{N} \lambda_k E_k(z) \ (N \to \infty) \\
E_k(z) &= c_k (R_{CSDA} - z) \cdot \exp[-\lambda_k (R_{CSDA} - z)]
\end{aligned}\right\} \qquad (74)$$

The aforementioned restriction to $N = 5$ is certainly extended to equation which can be considered as a representation of the BBE in terms of the residual energy $E(z)$. Due to the low-energy corrections ($a_0$, $a_{shell}$, $a_{Barkas}$) the energy-transfer function $dE(z)/dz$ remains finite for all z (i.e., $0 \leq z \leq R_{CSDA}$). This is, for instance, not true for the corresponding results according to Formulas (73 – 74) at $z = R_{CSDA}$. The calculation of E(z) and dE/dz according to equations, referred to as LET, is presented in Figure 5. The figure shows that, within the framework of CSDA, the LET of protons is rather small, except at the distal end of the proton track.

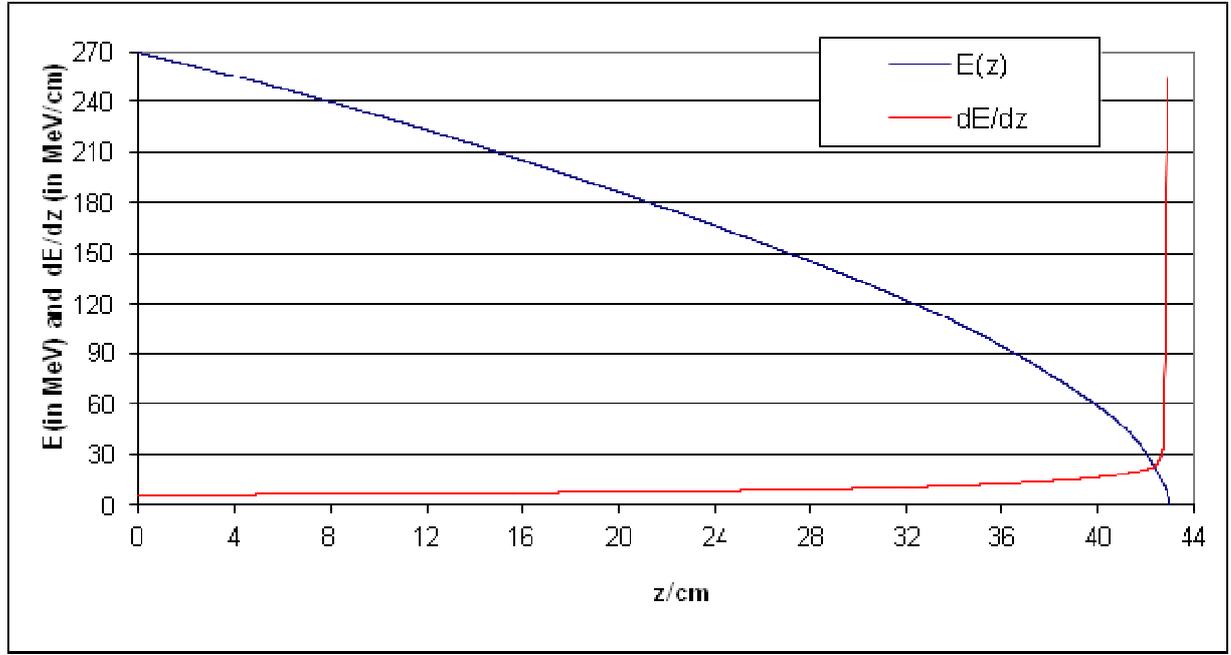

**Figure 5.** E(z) and dE(z)/dz as a function of z (LET based on CSDA); energy straggling is omitted.

A change from the interacting reference medium water to any other medium can be carried out by the calculation of $R_{CSDA}$, where the substitutions have to be performed:

$$R_{CSDA}(medium) = R_{CSDA}(water) \cdot (Z \cdot \rho / A_N)_{water} \cdot (A_N / Z \cdot \rho)_{medium}. \quad (75)$$

It is also possible to apply formula (75) in a stepwise manner (e.g., voxels of CT). This procedure will not be discussed here, since it requires a correspondence between $(Z \cdot \rho / A_N)_{Medium}$ and information provided by CT. With regard to heterogeneous media with only CT data as basis information the application of BBE is a more difficult task.

## 2.4 Qualitative properties of the electron transfer described by BBE and electron capture

According to BBE the energy spectrum of produced by carbon ions should be the same as that produced by protons, and the only difference between protons and carbon ions should be the intensity of the released collision electrons, i.e. the amplification factor should be 36 for carbon ions. It is well-known that this property is not valid for the following reasons: The average ionization energy for carbon ions turned out to be $E_I = 80$ eV instead of $E_I = 75$ eV for protons [41, 30], and [30] is based on investigations of some other authors [47, 56, 64, 65]. The second reason is the electron capture of the carbon ion. Thus a carbon ion can capture a free electron, which has been excited immediately before. Figure 7 shows this effect. However, only electrons with a slow relative velocity to the carbon ion can account for this process ($v_{relative}$ about 0). Since the transition time of the capture electron to a lower atomic state of the carbon ion is less than $10^{-10}$ sec with a simultaneous emission of light (UV or visible), it is possible that the captured electrons goes lost again, and only a stripping effect occurs for a short time. If the $C^{6+}$ ions has been finally transferred to a stable $C^{5+}$ ion, the identical process can be repeated until at the end track a neutral carbon atom is obtained having only a thermal energy. In the environment of the Bragg peak the effective charge of the carbon ion is about the same that of a proton, namely $+e_0$. Since the electron capture can only occur for electrons of which the relative velocity is slow, the upper energy limit of the energy exchange $E_{ex}$ is the Fermi edge $E_F$, which is for an electron gas not higher than the thermal energy $k_BT$. If the charge of carbon ion amounts to $+6 \cdot e_0$ and, at least, $> +e_0$, the environmental atomic electrons suffer lowering of the energy levels due to the Coulomb interaction, which leads to an increase of $E_I$. Therefore the stated value of $E_I$ = 80 eV represents an average value produced the fast carbon ion starting with $+6 \cdot e_0$ and ending with an uncharged, neutral carbon atom.

**Qualitative figure of projectile interaction of a charged particle**
**BBE:**

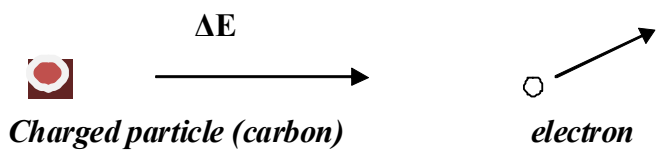

*Charged particle (carbon)*   *electron*

**Relative velocity between carbon and electron v = 0 (transition time < $10^{-10}$ sec) :**

**Electron capture by carbon ion**

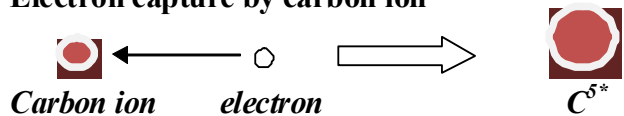

*Carbon ion*   *electron*   $C^{5*}$

**Figure 6.** Excitation of an atomic electron by the collision interaction of a fast carbon ion with an atomic electron and the reversal process of the electron capture.

*2.5. Boltzmann operator equation and Gaussian convolution*

In the following it is the task to obtain a quantum-statistical description of electron capture and stripping of electrons, i.e. those electrons which reduce the effective charge of the carbon ion for a short time and go lost before a transition to a stable atomic state of carbon can occur. For this purpose we consider the quantum statistical energy exchange $E_{ex}$ between projectile particle such as proton, He ion or carbon ion. The related mathematical procedure can be used to describe processes like energy straggling, lateral scatter and energy/charge exchange between projectile ion and released electrons below the Fermi edge $E_F$. However, before we can account for the latter problem we have to consider the related mathematical tools.

In general, if H represents the Hamiltonian (either non-relativistic or relativistic) and f(H) an operator functions, then for continuous operators H the connection holds:

$$\left. \begin{array}{l} H \cdot \Psi = E \cdot \Psi \\ f(H) \cdot \Psi = f(E) \cdot \Psi \end{array} \right\}. \quad (76)$$

At first we apply this relation in the non-relativistic case to derive the Gaussian convolution for the description of energy straggling. If the stopping power S(z) = dE(z)/dz of protons is calculated by BBE or by phenomenological equations [6] based on classical energy dissipation, then the energy fluctuations are usually accounted for by:

$$S(z) = \int S_{Rcsda}(u) K(\sigma, u-z) du. \quad (77)$$

This kernel may either be established by non-relativistic transport theory (Boltzmann equation) or, as we prefer here, by a quantum statistical derivation. Let φ be a distribution function and Φ a source function, mutually connected by the operator $F_H$ (operator notation of a canonical ensemble):

$$\left. \begin{array}{l} \varphi = \exp(-H/E_{ex})\Phi = F_H \Phi \\ F_H = \exp(-H/E_{ex}) \end{array} \right\}. \quad (78)$$

An exchange Hamiltonian H couples the source field Φ (proton fluence) with an environmental field φ by $F_H$, due to the interaction with electrons:

$$\left. \begin{array}{l} H = -\frac{\hbar^2}{2m} d^2/dz^2 \\ \exp(0.25 \sigma^2 d^2/dz^2)\Phi = F_H \Phi = \varphi \\ \sigma^2 = 2\hbar^2/mE_{ex} \end{array} \right\}. \quad (79)$$

It must be noted that the operator equation (79) was formally introduced [5] to obtain a Gaussian convolution as Green's function and to derive the inverse convolution. $F_H$ may formally be expanded in the same fashion as the usual exponential function $\exp(\xi)$; $\xi$ may either be a real or complex number. This expansion is referred to as Lie series of an operator function. Only in the thermal limit (equilibrium), can we write $E_{ex} = k_B T$, where $k_B$ is the Boltzmann constant and T is the temperature. This equation can be solved by the spectral theorem provided by the discipline *'functional analysis'*:

$$\left.\begin{array}{l} F_H \Phi = \gamma \Phi \\ \Phi_k = \exp(-ikz)/\sqrt{2\pi} \\ F_H \Phi_k = \gamma(k)\exp(ikz)/\sqrt{2\pi} = \exp(-\sigma^2 k^2/4)\cdot\exp(ikz)/\sqrt{2\pi} \\ K(\sigma, u-z) = \int \Phi_k^*(z)\Phi_k(u)\gamma(k)dk = \frac{1}{2\pi}\int \exp(-\sigma^2 k^2/4)\exp(ik(u-z))dk \\ K(\sigma, u-z) = \frac{1}{\sigma\sqrt{\pi}}\exp(-(u-z)^2/\sigma^2) \end{array}\right\}. \quad (80)$$

The kernel *K* according to equation (80) may either be established by non-relativistic transport theory (Boltzmann equation) or, as we prefer here, by a quantum statistical derivation. It is a noteworthy result [5, 6] that a quantum stochastic partition function leads to a Gaussian kernel as a Green's function, which results from a Boltzmann distribution function and a non-relativistic exchange Hamiltonian *H*. An operator formulation of a canonical ensemble is obtained by the following way: let $\phi$ be a distribution (or output/image) function and $\Phi$ a source function, which are mutually connected by the operator. In a 3D version, linear combinations of $K(\sigma, u - x)$ and the inverse kernel $K^{-1}$ are also used in scatter problems of photons [5]. As an example, we consider the Schrödinger equation of a free electron transferring energy from the projectile to the environment and obeying a Boltzmann distribution function $f(H) = \exp(-H/E_{ex})$:

$$H = -\frac{\hbar^2}{2m}\Delta. \quad (81)$$

The above relation provides:

$$\left.\begin{array}{l} \exp(-H/E_{ex})\cdot\exp(-i\vec{k}\cdot\vec{x})/(\sqrt{2\pi})^3 = \exp(-\hbar^2\vec{k}^2/(2mE_{ex}))\cdot\exp(-i(\vec{k}\cdot\vec{x}))/(\sqrt{2\pi})^3 \\ \vec{k}^2 = k_1^2 + k_2^2 + k_3^2 \end{array}\right\}. \quad (82)$$

In the case of thermal equilibrium, we can replace the exchange energy $E_{ex}$ by $k_B T$.

## *2.6. Dirac equation, Fermi-Dirac statistics and their consequences*

With regard to our task the Dirac equation to describe the particle motion is an adequate starting-point:

$$H_D = c\vec{\alpha}\cdot\vec{p} + \beta mc^2$$
$$\vec{\alpha} = \begin{pmatrix} 0 & \vec{\sigma} \\ \vec{\sigma} & 0 \end{pmatrix} \quad \beta = \begin{pmatrix} 1 & 0 \\ 0 & -1 \end{pmatrix} \quad . \quad (83)$$
$$H_D^2 = c^2 p^2 + m^2 c^4$$

3.

Please note that in the notation of equation (83) $\vec{\sigma}$ refers to the Pauli spin matrices (this should not be confused with the *rms*-value σ of a Gaussian distribution function). In position representation we obtain:

$$(\beta mc^2 + \frac{\hbar c}{i}\cdot\vec{\alpha}\cdot\nabla)\psi = E_D\cdot\psi \quad (84)$$

According [6, 63] we can write:

$$E_D = \pm mc^2\sqrt{1 + 2\cdot E_{Pauli}/mc^2}. \quad (85)$$

$E_{Pauli}$ is the related energy value resulting from the Pauli equation.

From the view-point of the many-particle-problem Fermi-Dirac statistics is adequate mean by the notation of operator functions:

$$f_F(\hat{H}) = \frac{1}{1+\exp[(H_D-E_F)/E_{ex}]}\cdot d_s(H_D). \quad (86)$$

$E_F$ represents the energy of the Fermi edge (usually some eV) and $d_s$ the density of states of the Hamiltonian $H_D$.

$$f_F(\hat{H})^n = [\frac{1}{1+\exp[(H_D-E_F)/E_{ex}]}\cdot d_s(H_D)]^n. \quad (87)$$

4.

We iterate equation (87) n-times and obtain:

$$f_F(\hat{H})^n = [\frac{\exp[-(H_D-E_F)/2E_{ex}]}{\exp[-(H_D-E_F)/2E_{ex}]+\exp[(H_D-E_F)/2E_{ex}]}\cdot d_s(H_D)]^n. \quad (88)$$

By that, the above expression assumes the shape:

$$f_F(\hat{H})^n = [\frac{1}{2} \frac{\exp[-(H_D - E_F)/2E_{ex}]}{\cosh[(H_D - E_F)/2E_{ex}]} \cdot d_s(H_D)]^n$$
$$= [\frac{1}{2} \exp[-(H_D - E_F)/2E_{ex}] \cdot \sec h[-(H_D - E_F)/2E_{ex}] \cdot d_s(H_D)]^n \quad (89)$$

Since $1/\cosh(\xi) = \text{sech}(\xi)$ holds, we evaluate equation (89) using an expansion resulting from Euler numbers $E_l$ [35]. Convergence is only established for $\xi \leq \pi/2$. Therefore we have derived a modified expansion which provides convergence for arbitrary arguments of $\xi$ [6]:

$$\sec h(\xi) = \exp(-\xi^2) \cdot \sum_{l=0}^{\infty} \alpha_{2l} \cdot \xi^{2l}$$
$$\alpha_{2l} = E_{2l}/(2l)! + \sum_{l'=1}^{l} (-1)^{l'+1} \cdot \alpha_{2l-2l'}/l'! \quad (90)$$

The spectral theorem of functional analysis provides:

$$\eta(k) = [mc^2\sqrt{1 + \hbar^2 \cdot k^2/m^2c^2} - E_F]/E_{ex}$$
$$\gamma(\eta(k)) = [\frac{1}{2}(\eta \cdot E_{ex} + E_F)/mc^2]^n \cdot \exp(-n\eta/2) \cdot \sec h(\eta/2)^n \quad (91)$$

By performing all integrations we obtain the distribution functions in the energy space (equation (91)) and position space (equation (91a)):

$$S_E = N_f \cdot \exp(-(E_n(k) - E_{Average,n})^2/2\sigma_E(n)^2) \cdot \sum_{l=0}^{\infty} b_l(n, mc^2) \cdot (E_n(k)/2E_{ex})^l. \quad (91a)$$

$$K_F = N_f \cdot \sum_{l=0}^{\infty} H_l((u - z - z_{shift}(l))/\sigma_n) \cdot B_l(n, mc^2) \cdot \exp(-(u - z - z_{shift}(l))^2/2\sigma_n^2) \quad (91b)$$

According to Bohr's formalism [38] the formula for energy straggling (or fluctuation) $S_F$ is given by:

$$S_F = \frac{1}{\sqrt{\pi}\sigma_E} \exp[-(E - E_{Average})^2 / \sigma_E^2] \quad (92)$$

The fluctuation parameter $\sigma_E$ can be best determined using the method in [38]. Furthermore we can verify the connection between $E_{Average}$ in the theory of Bohr and the Fermi edge energy $E_F$, since $E_{Average}$ results from the repeated iteration of $E_F$.

$$\left.\begin{aligned}
\Delta\sigma_E^2 &= \Delta z \cdot \tfrac{1}{2} \cdot (Z/A_N) \cdot \rho \cdot f \cdot \tfrac{2mc^2}{1-\beta^2}(1-\beta^2/2) \quad \text{(for finite intervals } \Delta z\text{)} \\
f &= 0.1535 \; MeVcm^2/g \\
d\sigma_E^2/dz &= \tfrac{1}{2} \cdot (Z/A_N) \cdot \rho \cdot f \cdot \tfrac{2mc^2}{1-\beta^2}(1-\beta^2/2)
\end{aligned}\right\} . \quad (93)$$

$\Delta\sigma_E^2$ contains as a factor the important magnitude $E_{max}$, that is, the maximum energy transfer from the proton to an environmental electron; it is given by $E_{max} = 2mv^2/(1-\beta^2)$. In a non-relativistic approach, we get $E_{max} = 2mv^2$. $E_{max}$ can be represented in terms of the energy E, and, for the integrations to be performed, we recall the relation $E = E(z)$ according to formula (93):

$$E_{max}(in \; keV) = \sum_{k=1}^{4} s_k \cdot E^k \quad (E \; in \; MeV). \quad (94)$$

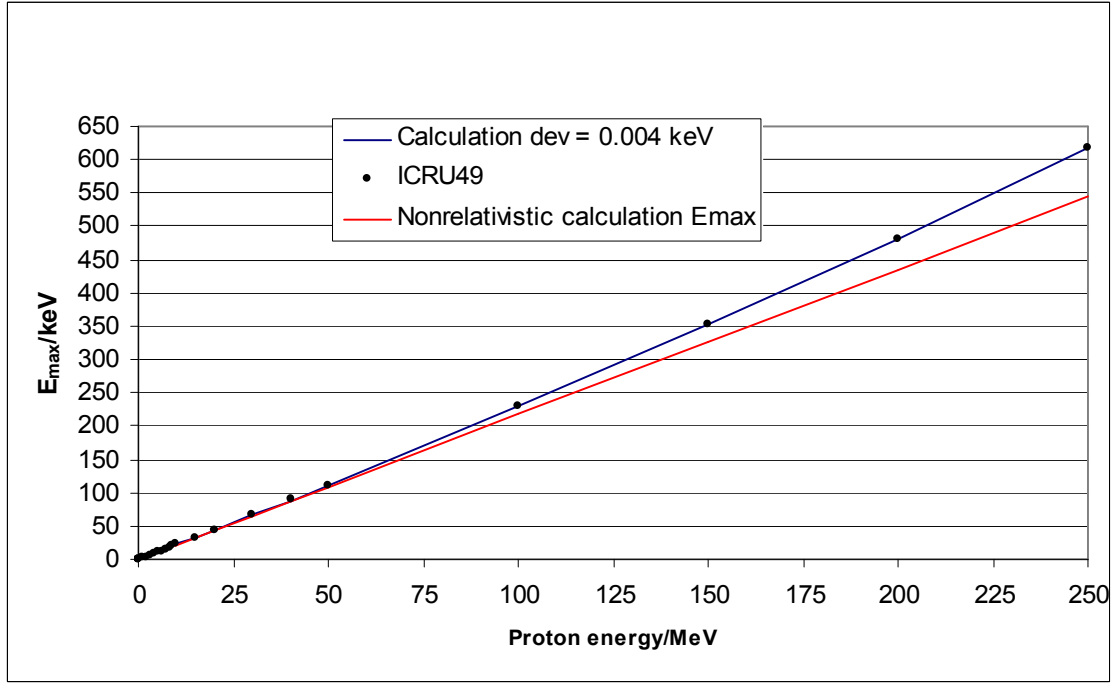

**Figure 7.** Calculation of $E_{max}$ according to equation (94). The straight line refers to the non-relativistic limit.

However, we should like to point out that according to the preceding section this determination is only valid for protons and cannot be applied to heavy ions without a change of the parameters.

**Table 5.** The parameters $s_k$ for the calculation of $E_{max}$ (formula (94)).

| $s_1$ | $s_2$ | $s_3$ | $s_4$ |
|---|---|---|---|
| 2.176519870758 | 0.001175000049 | -0.000000045000 | 0.0000000000348 |

As in a previous section we use the definition $S(z) = dE(z)/dz$ according to BBE. Since $S(z)$ is proportional to $q^2$, the following equation (95) provides $q^2(E) = q_0^2 \cdot S_E$.

$$[f_F(\hat{H})]^n S(z) = [\tfrac{1}{2}\exp[-(H_D - E_F)/2E_{ex}] \cdot$$
$$\cdot \sech[-(H_D - E_F)/2E_{ex}] \cdot d_s(H_D)]^n S(z). \qquad (95)$$

The transition to the integration (continuum approach up to second order) provides:

$$\left. \begin{array}{l} q^2(E) = q_0^2 [erf(E/s_E) - A \cdot (1 - E/E_0) \cdot (1 - (E/E_0) \cdot \exp(-E^2/s_E^2)] \\ s_E^2 = q_0^3 \cdot \pi \cdot m^2 c^4 (1 + 1/M^2 c^4) \\ A = q_0^2 \cdot \pi^2 \cdot mc^2 / Mc^2 \end{array} \right\} \qquad (96)$$

An essential result is that we are able to modify the previous formula between initial energy $E_0$ and the

range $R_{csda}$:

$$Rcsda = \alpha(E_0 \cdot N / q_{eff}^2) + \beta(E_0 \cdot N / q_{eff}^2)^2 + \gamma(E_0 \cdot N / q_{eff}^2)^3 + \delta(E_0 \cdot N / q_{eff}^2)^4$$
$$E_0 : initial\ \ energy\ /\ per\ nucleon\ ;\ N : nucleon\ \ number \quad (97)$$

Please note that the parameters have slightly to be modified $\alpha = 0.0069465598$; $\beta = 0.0008132157$; $\gamma = -0.00000121069$; $\delta = 0.000000001051$.

If $N = 1$ and $q_{eff} = 0.995$ the above formula is valid for protons. However, it turns out that the determination of the effective charge $q_{eff}$ depends on the initial energy $E_0$. This will be verified in a following section.

The subsequent Figure 8 indicates the wide tool resulting from linear combinations of shifted Gaussian kernels (the signs of the coefficient $c_1$ may considerably change, but $K_g > 0$ must still hold). Equations (91 – 91b) can be approximated by a linear combination of three kernels with different shift values and *rms* –values. Then it is possible to subject the corresponding deconvolutions based on the LNS-procedure to determine $q^2(E)$ from measured Bragg curves.

$$K_j = \frac{1}{s_j \sqrt{\pi}} \cdot \exp(-(u-x-x_j)^2 / s_j^2) \quad (j = 0, 1, 2)$$
$$c_0 + c_1 + c_2 = 1 \ (c_1 < 0)$$

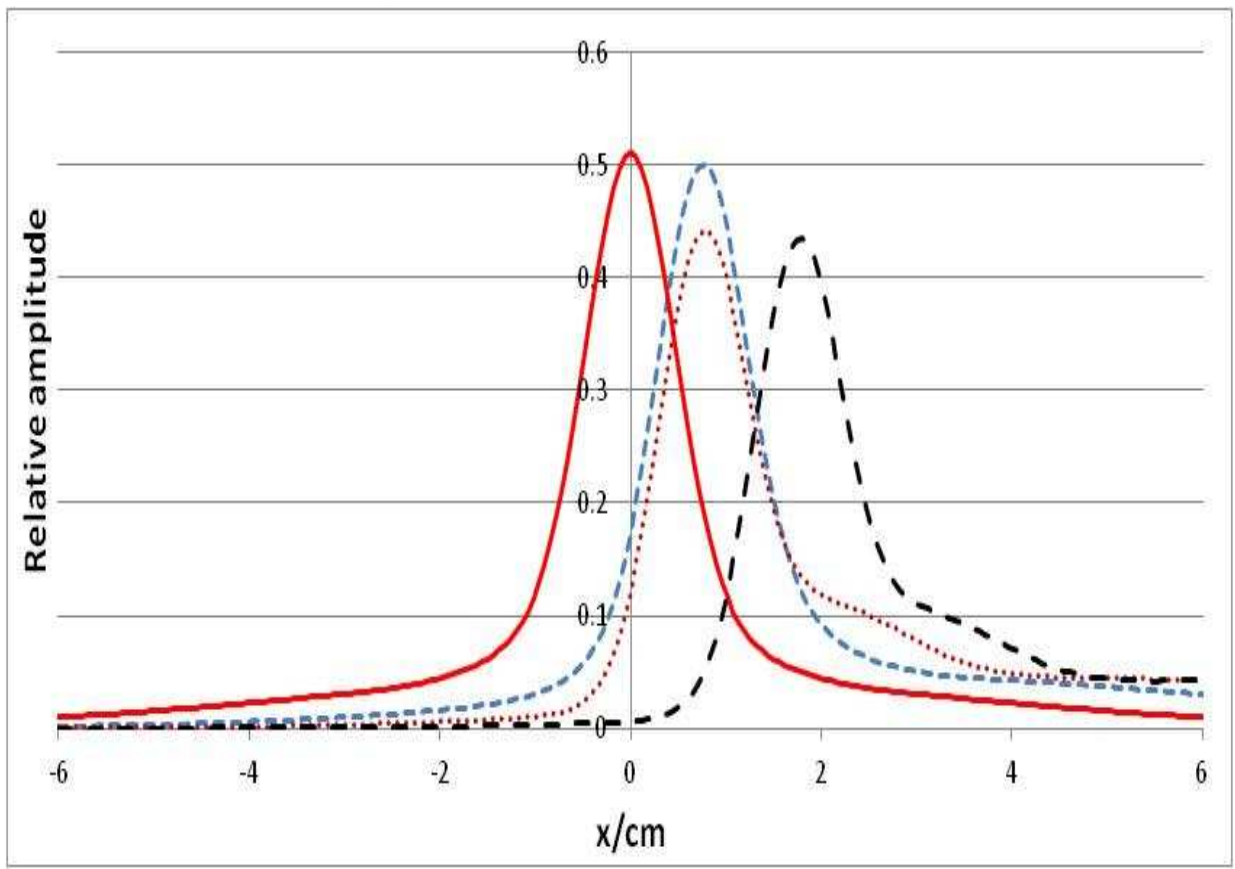

**Figure 8.** Linear combination of three Gaussian kernels with different shifts $x_j$ to describe asymmetric processes like electron capture or Landau tails in energy straggling.

*2.7. Monte-Carlo methods*

With regard to problems of image processing (e.g. blurred images due to scatter effects) we have carried out Monte-Carlo calculations with the EGSnrc code [31]. This code has been applied most widely to various tasks in medical radiation physics. We have performed Monte Carlo calculations using the EGSnrc code with regard to problems of image processing in the *MV- and KV-domain*. The transport of charged particles and the related nuclear reactions have been determined with the aid of GEANT4 [42].

*2.7.1. Image processing in* **the** *MV-domain*

Absorption – and attenuation curves, transverse profiles in various depths for the simulation of radiation responses of a detector array (portal imaging) have been determined for field sizes 0.48 x 0.48 cm$^2$ up to 20 x 20 cm$^2$. Previous results have been used with regard to the energy spectrum of 6 MV [7].

*2.7.2. Image processing in the KV-domain*

We have determined the energy spectrum of 100 KV and 125 KV photons of CT/CBCT and the absorption/scatter behavior in some media of relevance, e.g. water-equivalent and phantoms with different material densities (lung, bone). A main purpose was the connection between Hounsfield units and the scatter parameters required for the 2D scatter kernel:

$$K_g = c_0 \cdot K_0(s_0, u-x, v-y) + c_1 \cdot K_1(s_1, u-x, v-y). \tag{98}$$

In general, the scatter parameters $s_0$ and $s_1$ depend (increase) on the depth z, and this is the way to treat the depth-dependent scatter of a pencil beam. The correspondence between the Hounsfield value and electron density ρ is well-established, excepted metallic implants. The scaling of the scatter parameters $s_0$ and $s_1$ can be scaled according to the electron density ρ, if the scatter parameters are known for water. A possible, but rather intricate way to eliminate scatter in CT/CBCT images would be obtained by the deconvolution of photon pencil beams, i.e. the methods of radiation therapy planning [7] are transformed to image processing.

Therefore, we extend here a previously developed method of the deconvolution according to the volume [5] to the parallel solution procedure of LNS presented in this study.

*2.7.3. Extension of the LNS method to volume-dependent scatter functions s, $s_0$, $s_1$ and $s_2$*

We have already pointed out that the scatter parameters s, $s_0$, $s_1$ (and eventually $s_2$) have by no means to be constant values. Thus, in the pencil beam algorithms [6 – 8] these parameters are not constant, but they represent scatter functions depending on the depth z. However, this restriction is, in general, not necessary in all formulas we have developed in this study.

The differential operator formulations of one and/or more than one kernel expressed by $O^{-1}$, $O$, $O_g^{-1}$, $O_g$, permit a dependence of all parameters s, $s_0$, $s_1$, $s_2$ and related composite terms like $\sigma$, $\sigma_1$, $\sigma_2$ of all three dimension magnitudes x, y, z, since the differential operators in the exponential functions are not influenced by this property. This property is also true with regard to all integral operator formulations (including IFIE2 and LNS procedure), where the half-width parameters do not affect the integration variables. In all our applications, we do not account for neither complex-valued Gaussian kernel functions nor source/image functions ρ and φ. We restrict ourselves to positively definite source/image functions. Thus we have previously used Fourier expansions of the scatter functions [5], and the same procedures are now applied to some cases of the LNS calculations (image problems):

$$\left. \begin{array}{l} s_k(x,y,z) = \sum_{j1,j2,j3=0}^{\infty} A^k_{j1,j2,j3} \cdot \cos(\vec{j} \cdot \vec{x}) + B^k_{j1,j2,j3} \cdot \sin(\vec{j} \cdot \vec{x}) \\ k = 0, 1, 2 \end{array} \right\}. \qquad (99)$$

Equation (99) is of particular importance, if the source function ρ is connected to a dose distribution function D without scatter, i.e. it only contains absorption but not attenuation, whereas the image function φ represents a blurred dose distribution which also contains scatter.

### 2.8. *Measurement data and calculations via therapy planning system*

In this communication we have used the algorithm AAA [7] implemented in the planning system Eclipse[R] (Varian, installation in the Klinikum Frankfurt/Oder). The radiation leaving a phantom has been recorded with the Iview[R] (Synergy, Elekta). Details of CT/CBCT measurements have been previously given [5]. A stereotactic photon beam has been recorded with a Novalis accelerator (Varian); proton beam data have been made available from the Harvard cyclotron (HCL), Boston.

With regard to measurement data the problem of noise is not significant in high dose radiotherapy (protons, photons), since data fluctuations are extremely small, temperature and pressure corrections have to be accounted for, and the detected raw data are always refined by specific procedures of smoothening. Thus these data can readily compared to theoretical calculations. In the case of image processing carried out at lower doses (CT, CBCT) the situation is not quite as favorable as in radiotherapy, and smoothening plays a more significant role to remove noise produced by fluctuations in the detector system. These fluctuations may result from local temperature influences and/or the memory of detectors due to preceding signal sequences. However, these data are also refined by appropriate software, which is already accounted for by the vendors.

## 3. Applications

Since we are interested in calculation results and the reliability obtained by the LNS expansion, the following section accounts for those examples we have already discussed in detail by published methods [5]. For brevity, we select some cases; further examples can be obtained upon request. By that, the algorithms concerning the inverse problem of linear combinations of Gaussian convolution kernels will gain more flexibility. A further application is the formulation of electron capture by an asymmetric kernel based on Fermi-Dirac statistics.

### *3.1. Comparison of LNS procedure with a previously published method*

In order to check the reliability and convergence properties of the LNS procedure, we perform at first applications we have previously obtained by the calculation of $K_g^{-1}$. In both calculation procedures, we need the deconvolution kernel $K_0^{-1}(s_0, u-x)$, which has to be appropriately extended, if necessary, to more than one spatial dimension. Since $K_0^{-1}$ represents itself an infinite expansion, we denote here the finite break off value by *N* we have used in a corresponding calculation. It has to be pointed out that for a reasonable comparison of the two different inverse procedures N has to be identical in both cases. The finite break off value of the sequence of Gaussian convolution terms according to a previous study [5] will be denoted by *M*, and the related value of the LNS procedure according to equations (53 -62) by *L*. The best test of the derived deconvolution formulas can be obtained by corresponding convolutions of some model cases and back calculations via LNS procedure. Since the deconvolution formulas represent order-by-order calculations, a principal aim of the tests was to specify the required order and precision to obtain the source function (origin) in a satisfactory way.

The principal problems of deconvolutions and possible pitfalls can be verified either by the Figures 9 – 10 or by Figures 15 - 17 in section 3.2. These figures show that rather different sources (e.g. three adjacent boxes or boxes with an empty space between them) with different *rms* values s, $s_0$, $s_1$ and $s_2$ lead to similar image functions. By that, we have to verify that the underlying *rms* values have to be known rather exactly from measurement data or by Monte-Carlo calculations to prevent artifacts by the deconvolution procedures. Only due to the very accurate knowledge of the subjected convolution parameters it is possible that the inverse procedure also reliable works with sufficient accuracy. The so-called 'try-and-error' method with certain start values for the *rms* parameters s, $s_0$, $s_1$, $s_2$ and coefficients $c_0$, $c_1$, $c_2$ might lead to artifacts. The model cases according to Figures 9 – 10 and 15 – 17 may also have a practical importance, since the boxes represent finite step functions, where the $L_1$-integrability is rather favorable to handle, and the deconvolution via Fourier transforms and Wiener filters leads to diverging jumps at the edges (this is a typical ill-posed problem [22]). In radiotherapy the fluence determination/modulation and optimization within finite intervals (grid size) represents an important feature in IMRT (or Rapid Arc) therapy. On the other side, only by rescaling of the underlying geometry we are directly guided to these aspects and novel treatment schemes of modern radiotherapy [23], which appear to lead to a better protection of critical organs and to fulfill the corresponding constraints of radiobiology and radiation protection.

**Table 5.** Convolution/deconvolution parameters of Figures 9 – 10.

| Figures | $c_0$ | $c_1$ | $c_2$ | $s_0$/cm | $s_1$/cm | $s_2$/cm | L | M | N |
|---|---|---|---|---|---|---|---|---|---|
| 9 | 0.90 | -0.38 | 0.48 | 0.40 | 0.82 | 1.50 | 11 | 10 | 10 |
| 10 | 0.90 | -0.38 | 0.48 | 0.44 | 0.78 | 1.55 | 10 | 11 | 10 |

The parameters of Table 5 refer to that case, where we have accounted for a linear combination of three Gaussian convolution kernels, but with $c_0$ and $c_2 > 0$ and $c_1 < 0$ ($c_0 + c_1 + c_2 = 1$). This case is also supported by the LNS procedure and increases the flexibility of convolution/deconvolution applications without having to consider the Mexican hat problem with $c_1 < 0$ and $c_2 = 0$. We should recall that in spite of the modification with $c_1 < 0$ the condition $K_g \geq 0$ has to be satisfied. This property certainly represents a constraint at the choice of $c_1$ and $s_1$.

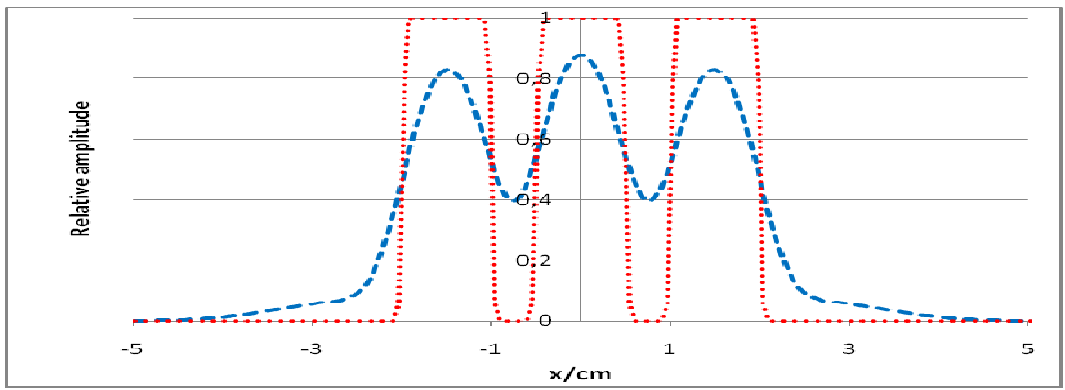

**Figure 9.** Convolution and deconvolution of three boxes (box length: 1 cm, space between them: 0.5 cm).

In contrast to Figure 9 the distance between the three boxes is increased in Figure 10; the relative amplitude between the boxes obtained after convolution reflects this property.

This property represents an essential restriction with regard to the choice of the scatter parameters and coefficients of the linear combinations $c_0$, $c_1$ and $c_2$. The negative value of $c_1$ yields the rapid decrease of the relative amplitude at the outermost boxes. The relative amplitude is not yet specified; it might refer to a fluence or dose distribution or to a signal strength in some other kinds of applications, where convolutions and their inverse problems are applied (e.g. image processing based on magnetic resonance tomography).

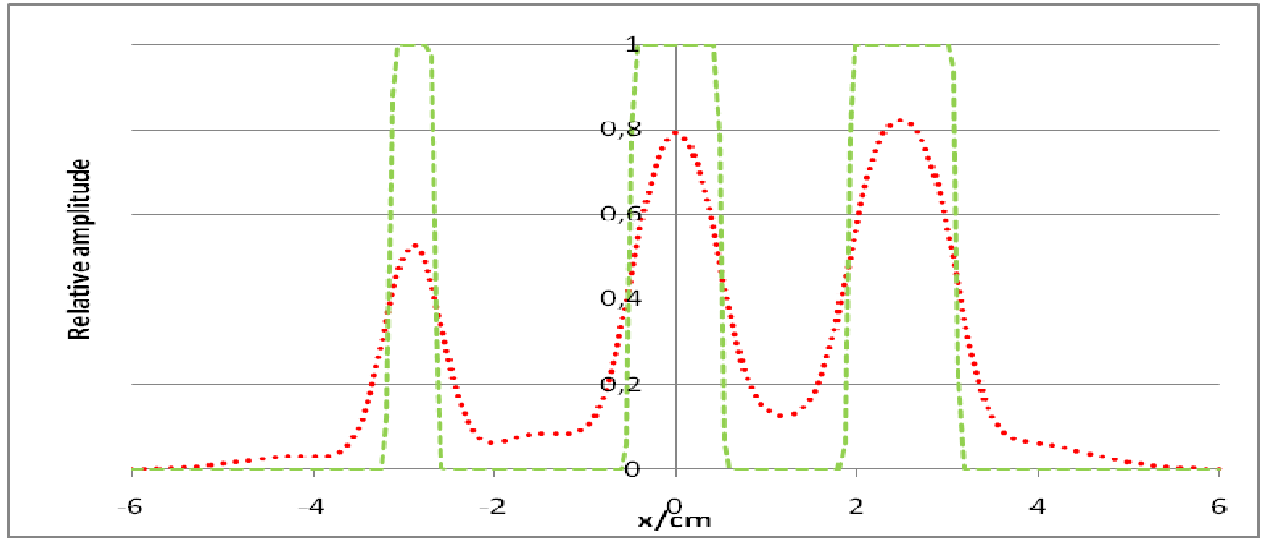

**Figure 10.** Convolution/deconvolution of three boxes (box lengths: 1 cm (right-hand side), 0.8 cm (central part), 0.5 cm (left-hand side); spaces: 1.6 cm (right-hand side), 2.35 cm (left-hand side)).

The following examples (Figures 11 – 15) represent a modification of a previously considered image deconvolution [5] of a test phantom (CT image), whereas we now consider the same phantom configuration in connection with a CBCT image. The test phantom (Figure 11) consists of an inner cylinder with a diameter of 4 cm (HU = 700) embedded by an outer tube containing water-equivalent material (HU = 0); the total phantom diameter amounts to 16 cm. The impinging photon beam with 140 KV now is a broad beam (CBCT), which can be calculated from a Gaussian beamlet with $s_0$ = 0.87 mm, whereas in the previous study we have used a Gaussian beamlet with 125 KV (scanning technique with CT, $s_0$ = 0.5 mm). Please note that the removal of noise of detectors recording images ('raw data') has been performed by a smoothing function established in the algorithm (Figures 11 – 15, 19 – 24). The application of a deconvolution procedure as previously used [5] or in this study via LNS requires the determination of the scatter in the phantom. The energy spectrum of the incident photon beam has been determined by Monte-Carlo calculations [31]. In both cases, CT and CBCT, a detector array records the attenuation radiation behind the phantom. The valid scatter functions for the CT imaging have already been presented [5], and the proportionality between the electron density $\rho_{el}(x, y, z)$ and the scatter functions $s_0(x, y, z)$, $s_1(x, y, z)$, and $s_2(x, y, z)$ holds ($c_0$, $c_1$, $c_2$ remain unchanged). However, the end values of the scatter functions at the detector plane are not valid. The scaling transformation has to be corrected by the detector influence and the initial scatter of the photon beam at the impinging position:

$$s_0(x, y, z) = s_{0,i} \cdot \rho_{el}(x, y, z) \, . \qquad (100)$$
$$s_1(x, y, z) = s_{1i} \cdot \rho_{el}(x, y, z). \qquad (100\ a)$$
$$s_2(x, y, z) = s_{2,i} \cdot \rho_{el}(x, y, z) \, . \qquad (100\ b)$$
$$s_{0i} = 0.42 \ cm \ ; \ s_{1i} = 0.56 \ ; \ s_{2i} = 2.72 \ cm \, . \qquad (100\ c)$$
$$c_0 = 0.87 \ ; \ c_1 = -0.12 \ ; \ c_2 = 0.25 \, . \qquad (100\ d)$$

The number of linear combinations of kernels (two kernels for CT and three kernels for CBCT) is the principle difference between the parameters according equations (100 – 100d) valid for CBCT and those parameters valid for CT. Moreover, $s_{1i}$ does not satisfy $s_{1i} > \sqrt{2} \cdot s_{0i}$; therefore the use of the previous algorithm is difficult to handle; a detailed treatment of this situation has been previously given [5]. The uncorrected electron density functions result from the Fourier expansion (99), where the scatter influence is accounted for. We have now to perform the task that the deconvolution of the 3D image at the central ray should provide the same result as one image of CT.

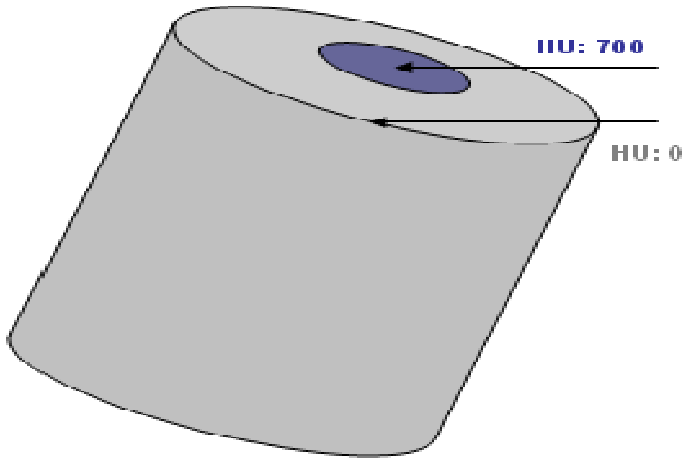

**Figure 11.** Phantom: water-equivalent material/bone. Inner cylinder: bone with HU = 700, outer part: water-equivalent material with HU = 0. In a later application the bone material with HU = 700 will be replaced by air and serves as a test phantom for a portal imager of a linear accelerator.

In the case of CT image processing, the cylinder is scanned along the cylinder axis, whereas in CBCT image processing the image is produced via one rotation by divergent broad beam. The problem of scanning by CBCT is certainly an increased contribution of scatter by the X-rays. This is a characteristic feature of all cases of broad beams and not only restricted to the KV domain. The deconvolution problem of the cylinder (Figure 11) based on CT scanning has been previously reported [5].

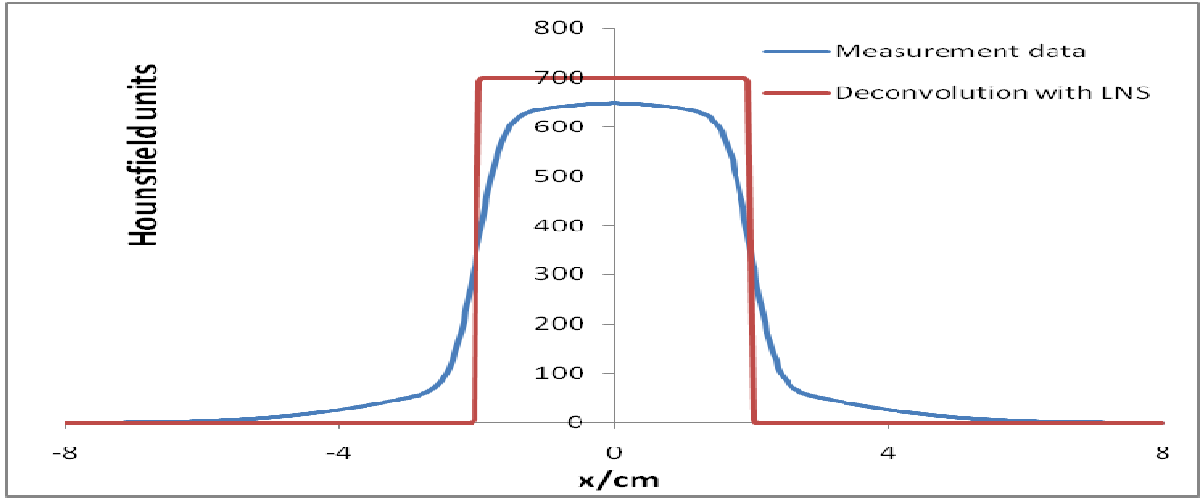

**Figure 12.** Profile of the Hounsfield units (CBCT) of the phantom cylinder (N = 7, L = 8).

Equations (100 – 100d), which determines the space-depending scatter function useful for deconvolutions of the complete volume, have to be modified in the CBCT case:

$$s_0(x,y,z) = s_{0,i} \cdot \rho_{el}(x,y,z) \cdot Cf. \qquad (100e)$$
$$s_1(x,y,z) = s_{1i} \cdot \rho_{el}(x,y,z) \cdot Cf. \qquad (100f)$$
$$s_2(x,y,z) = s_{2,i} \cdot \rho_{el}(x,y,z) \cdot Cf. \qquad (100g)$$

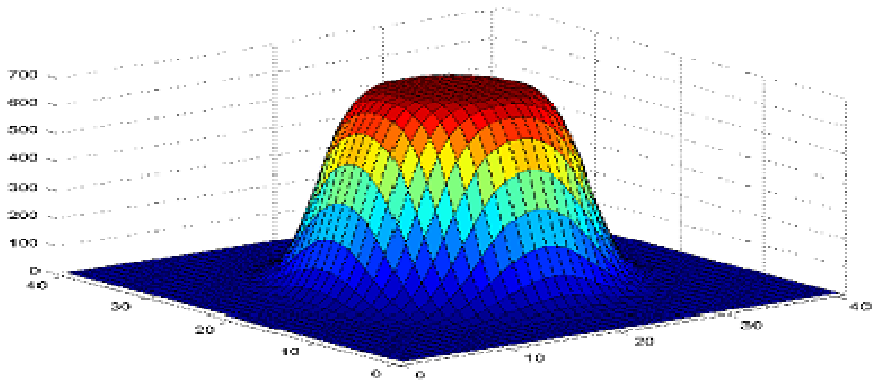

**Figure 13.** Hounsfield units of a 2D cylinder based on a measurement with a detector array.

The factor *Cf* results from the divergence of the X-rays. Only in the central ray we have to put *Cf = 1*.

We have to point out that it is a feature of CT that divergent rays are not used. However, in our case of CBCT application with rotational symmetry the factor *Cf* is determined by

$$Cf = \sqrt{(d^2 + SAD^2)/SAD^2} \ . \tag{101}$$

In equation (101) SAD refers to the source-axis-distance and d to distance from the center of the central axis of the cylinder and its rotation axis.

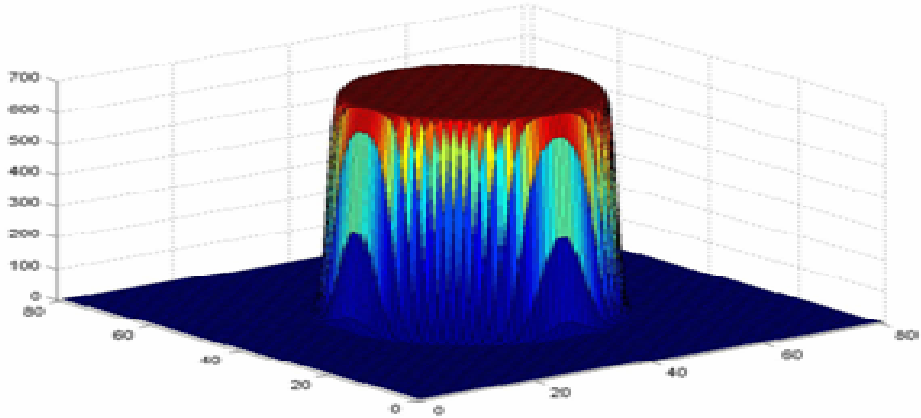

**Figure 14.** Hounsfield units of a 2D cylinder (Figure 14 represents the result after deconvolutions of Figure 13).

Thus Figure 14 is chosen such that the central ray of CBCT scanning is identical with the result of CT scanning. It is obvious that we need significantly more effort in the CBCT case with regard to the inverse problem, namely the order L of the deconvolution procedure. On the other side, this scanning technique provides a complete 3D image. In order to obtain reliable results in the CBCT case, the calculations had to be performed by accounting for higher order terms in the LNS procedure. In this communication, we have only considered the inverse problem of the central ray, but with regard to CBCT the calculation procedure of the inverse problem requires the modifications according to equation (101).

### 3.2. *Further results obtained by LNS in image processing and proton/photon dosimetry*

The examples presented in Figures 10 – 12 may serve as further tests of inverse calculations via LNS with possible applications to IMRT/IGRT. A comparison of Figure 15 with Figures 16 – 17 demonstrates the possible pitfalls of deconvolutions. Thus it is clear that the convolution of a triangle provides a 'triangle' with rounded corners. However, the shape of the images obtained via convolution of non-adjacent boxes might lead to the assumption that the source function has also the shape of a triangle, which is apparently not true. This fact clearly demonstrates that 'try-and-error' methods to determine the parameters for the inverse procedures might lead to artifacts.

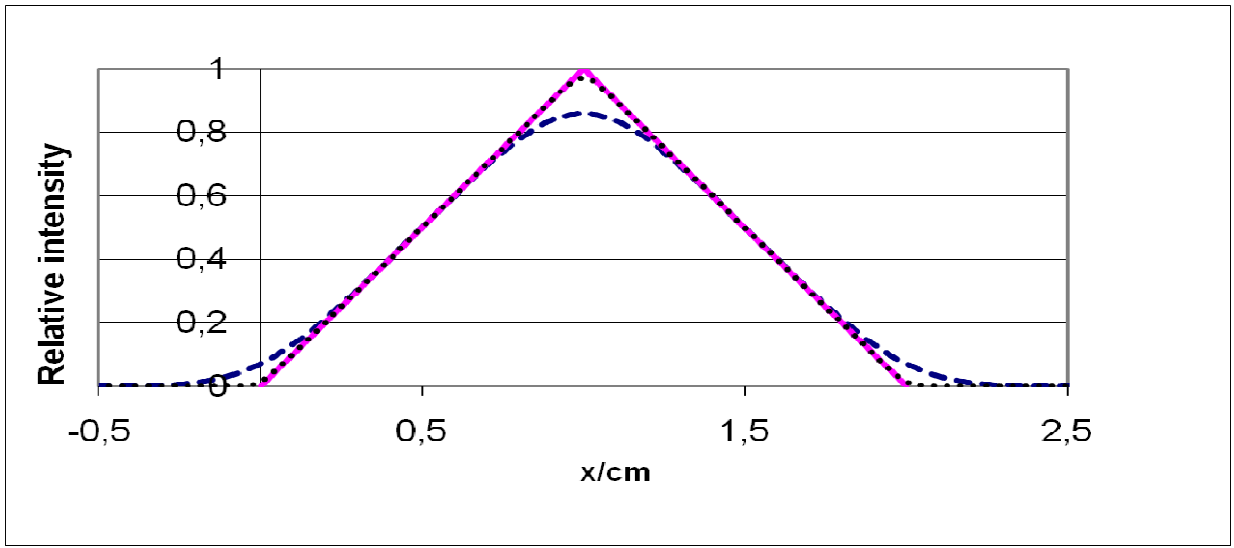

**Figure 15.** Convolution of a triangle (solid) with one Gaussian kernel (dashes) and deconvolution (dots).

**Table 6.** Convolution/deconvolution parameters in Figures 15 – 17.

| Figure | $c_0$ | $c_1$ | $s_0$/cm | $s_1$/cm | L | M | N |
|---|---|---|---|---|---|---|---|
| 15 | 1 | - | 0.25 | - | - | - | 4 |
| 16/dashes | 1 | - | 0.10 | - | - | - | 12 |
| 16/dots | 0.80 | 0.20 | 0.025 | 0.075 | 15 | 15 | 12 |
| 17/solid | 1 | - | 0.09 | - | - | - | 12 |
| 17/dashes | 0.80 | 0.20 | 0.015 | 0.050 | 15 | 16 | 12 |

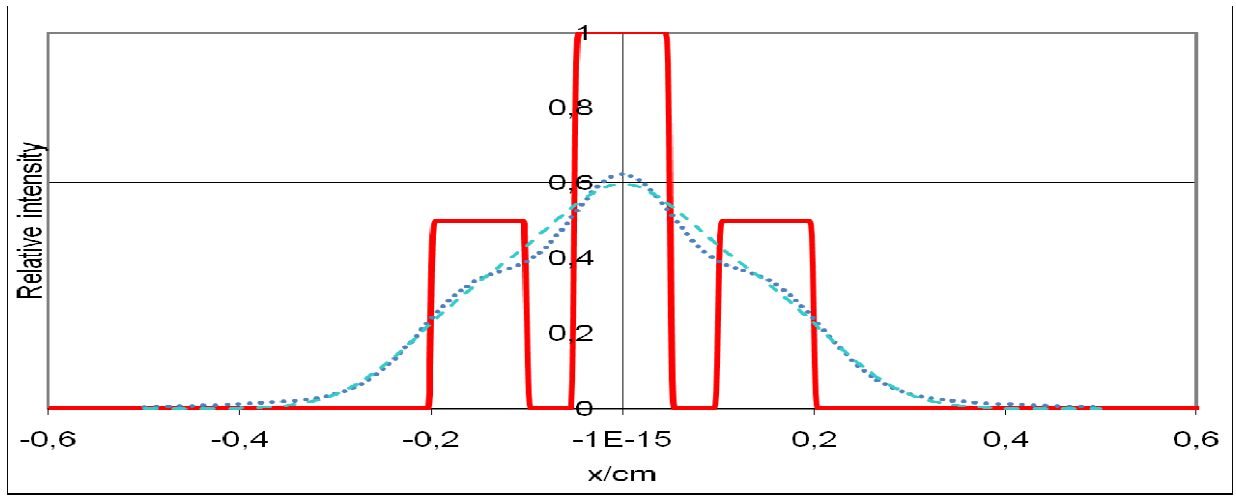

**Figure 16.** Convolution of three boxes (box length: 0.1 cm, space length between the boxes: 0.05 cm, height of the source functions: 1 (middle part) and 0.5 (at both sides)). Deconvolution: identical with the solid boxes, rounded corners not verifiable.

The deconvolution procedure by the LNS method has been applied (L = 20, N = 20) in Figure 16. The reason for the increased effort results from the long-range tail of the scatter of the high energy bremsstrahlung.

Further applications with photon beams are shown in Figures 16 – 15. Instead of image creation with X-rays (CT, CBCT) a portal imager (6 MV, bremsstrahlung) has been applied. Due to the long range of lateral scatter the portal imager does not provide the same height of the central ray as is can be verified from the previous figures (KV domain), and the lateral tail has significantly been increased. The deconvolution procedure has also to be performed by accounting much more terms of higher order in the LNS procedure than in the previous cases, and some noteworthy roundness can be verified in spite of the increased effort with regard to the order L.

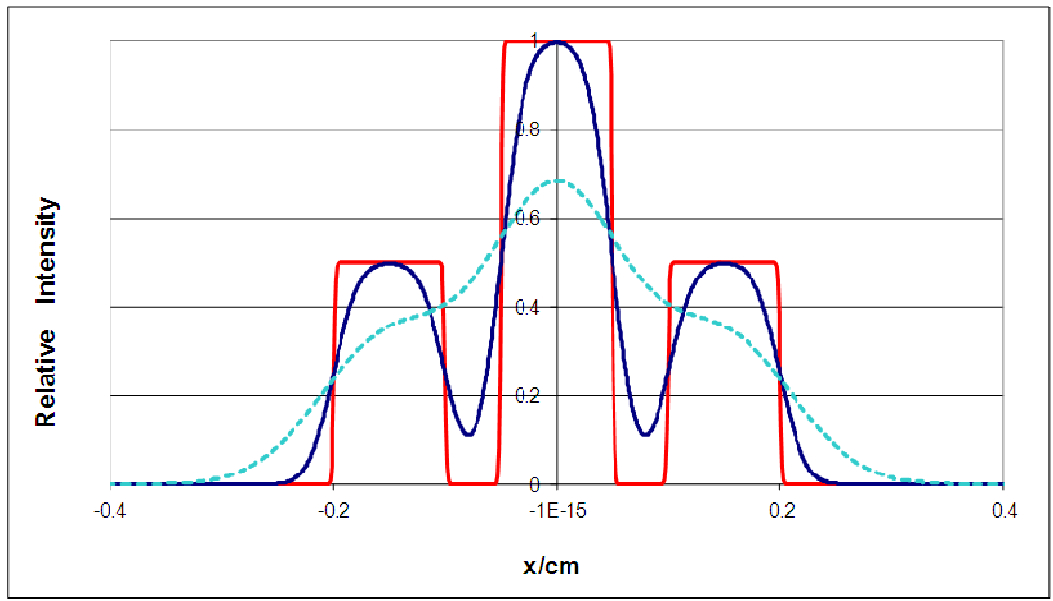

**Figure 17.** Geometry and box heights: see Figure 16. Convolutions have been obtained with different parameters. Deconvolutions are considered as identical with the origin (solid boxes), if the rounded corners are not verifiable.

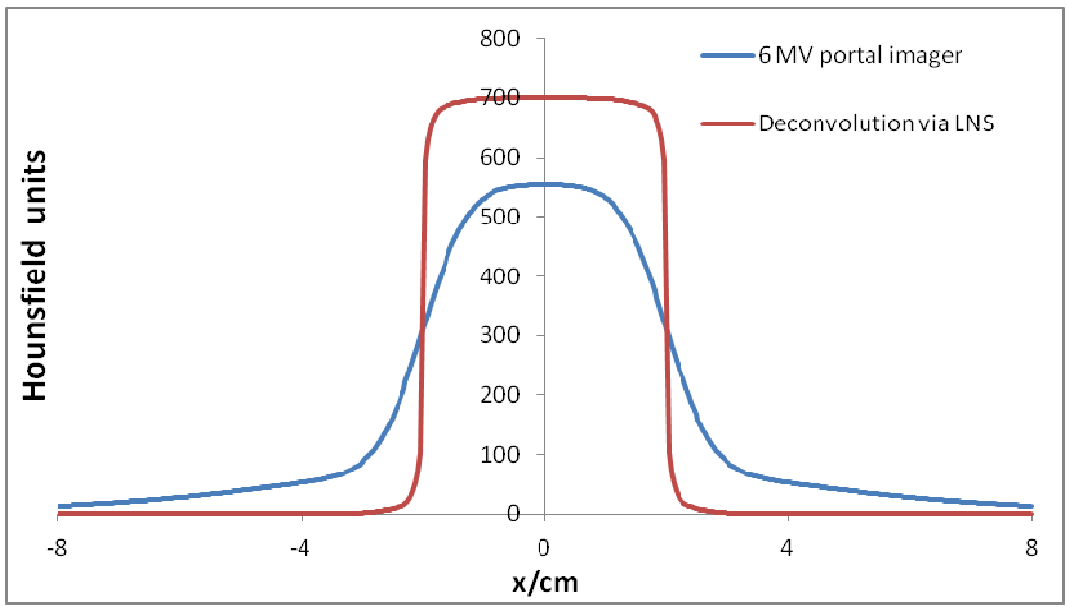

**Figure18.** Water/air phantom as a modified configuration of Figure 11 and image produced by 6 MV bremsstrahlung in a portal imager system (L = 11, N = 10).

The *rms* parameters for the application of the LNS method have been used from a previous publication [7]:

$$c_0 = 0.66; \; c_1 = 0.26; \; c_2 = 0.08. \tag{102}$$

$$s_0 = 0.82 \text{ cm}; \; s_1 = 4.735 \text{ cm}; \; s_2 = 12.334 \text{ cm}. \tag{102a}$$

The calculation with the previous method (M = 29, N = 20) revealed a superiority of the LNS procedure to provide faster convergence in the case of long-rang tails.

We have also used a modified configuration of the cylinder according to Figure 11, namely with air in the inner part instead of bone-equivalent material, for a further measurement with a portal imager (6 MV, bremsstrahlung). Based on the LNS method the measurement data have been analyzed with parameters of the publication cited above [7]:

$$c_0 = 0.66; \; c_1 = 0.26; \; c_2 = 0.08. \tag{103}$$

$$s_0 = 0.39 \text{ cm}; \; s_1 = 1.77 \text{ cm}; \; s_2 = 6.16 \text{ cm}. \tag{103a}$$

Figures 18 - 19 show the adaptation of the measurement data with the help of the AAA algorithm and the deconvolution via LNS (boxes with weak roundness at the corners). In contrast to Figure 18, where we have used Hounsfield units as the reference scale, we present in Figure 19 the density (cross-section).

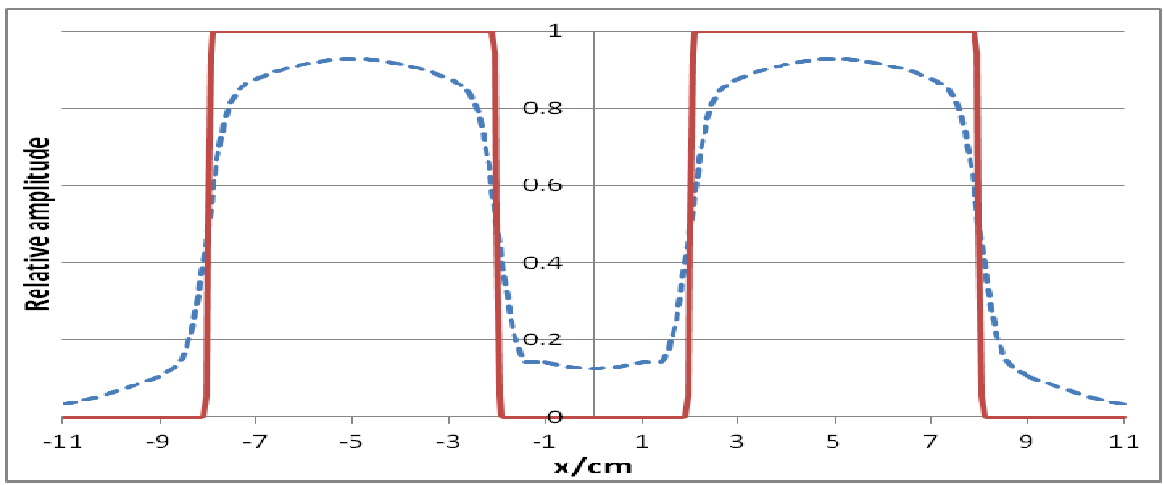

**Figure 19.** Water/air phantom as a modified configuration of Figure 11 and image produced by 6 MV bremsstrahlung in a portal imager system (L = 11, N = 10).

### 3.3. *Applications of LNS to photon/proton dosimetry (Stereotaxy and IMPT)*

Processing of very small field-sizes is a pathological situation in photon and proton dosimetry, since modern irradiation techniques such as proton beam scanning (with and without intensity modulation) and Stereotaxy/RapidArc require the handling of extremely small fields. In order to reach a comparable situation, we use both for proton and photon beam 0.48 x 0.48 cm² field size. With regard to depth dose/fluence decrease we have to distinguish between pure energy absorption and attenuation of a beam or a simple beamlet (the latter case also accounts for the influence of scatter). Figure 21 presents both absorption and attenuation. In particular, very small field-sizes have a significant influence to attenuation due to scatter; the related curves are rather different. The fluence decrease curve can be described as follows:

$$\Phi = \Phi_0 \cdot [a_0 \cdot \exp(-\mu_0 \cdot z) + a_1 \cdot \exp(-\mu_1 \cdot z)] \cdot \left[ \begin{array}{l} c_0 \cdot (erf(\tfrac{a-x}{s_0}) + erf(\tfrac{a+x}{s_0})) \cdot (erf(\tfrac{a-y}{s_0}) + erf(\tfrac{a+y}{s_0})) \\ + c_1 \cdot (erf(\tfrac{a-x}{s_1}) + erf(\tfrac{a+x}{s_1})) \cdot (erf(\tfrac{a-y}{s_1}) + erf(\tfrac{a+y}{s_1})) + c_2 \cdot (erf(\tfrac{a-x}{s_2}) + erf(\tfrac{a+x}{s_2})) \cdot (erf(\tfrac{a-y}{s_2}) + erf(\tfrac{a+y}{s_2})) \end{array} \right] . \quad (104)$$

$$s_0 = s_{00} \cdot z; \qquad s_1 = s_{10} \cdot z; \qquad s_2 = s_{20} \cdot z \quad (104a)$$

**Table 7.** Parameters of formula (104) and the scatter functions (104a).

| $a_0$ | $a_1$ | $\mu_0[\text{cm}^{-1}]$ | $\mu_1[\text{cm}^{-1}]$ | $c_0$ | $c_1$ | $c_2$ | $s_{00}[\text{cm}]$ | $s_{10}[\text{cm}]$ | $s_{20}[\text{cm}]$ |
|---|---|---|---|---|---|---|---|---|---|
| 0.748 | 0.252 | 0.01502 | 0.02204 | 0.605 | 0.246 | 0.149 | 0.035 | 0.210 | 0.525 |

Formulas (104) and (104a) describe a pencil beam model of 15 MV photons. The deconvolution procedure can be carried out with both methods presented in this communication. In order to obtain the pure absorption curve via measurement data, the influence of diamond detector had to be removed by an additional deconvolution. The refined measurement data agree with the theoretical model [7], if the scatter functions $s_0(z)$, $s_1(z)$, $s_2(z)$ are subjected to a deconvolution with σ = 1 mm to account for the finite size of the detector and its additional scatter influences. Therefore we have to perform the substitutions:

$$s_0' = \sqrt{s_0^2 + \sigma^2}; \quad s_1' = \sqrt{s_1^2 + \sigma^2}; \quad s_2' = \sqrt{s_2^2 + \sigma^2}. \quad (104b)$$

The deconvolutions have to be carried out using the corrected scatter functions $s_0'$, $s_1'$ and $s_2'$.

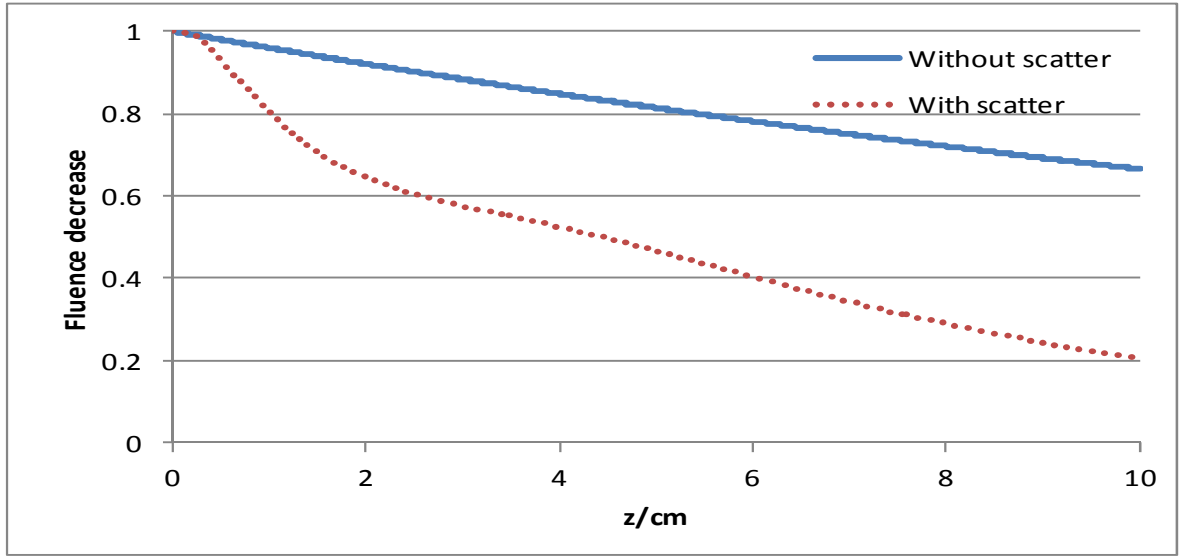

**Figure 20.** Fluence decrease of 6 MV bremsstrahlung, field-size: 0.48x0.48 cm$^2$ (solid line: without scatter (pure absorption), dashed line: with scatter (attenuation)).

The resulting attenuation curve and the lateral profiles are presented in Figures 20 and 22, which indicate the role of detectors and usual photon scatter in very small field-sizes. It should be mentioned that in IMRT and Stereotaxy we have very often to deal with photon beams with this order of magnitude. The deconcolution procedures applied to Figures 20 – 25 have been performed with L = 12, M = 17 and N = 15 to reach identical results. The superiority of LNS can be recognized again.

In order to obtain the real absorption profiles according to Figure 21 and 22 the deconvolution of three Gaussian kernels have to be performed. This can be done with the previous method and with the help of LNS procedure. The former method is applicable, since the convergence criterion can be satisfied, i.e. $s_o(z) < 2^{0.5} \cdot s_1(z)$ and $s_o(z) < 2^{0.5} \cdot s_2(z)$.

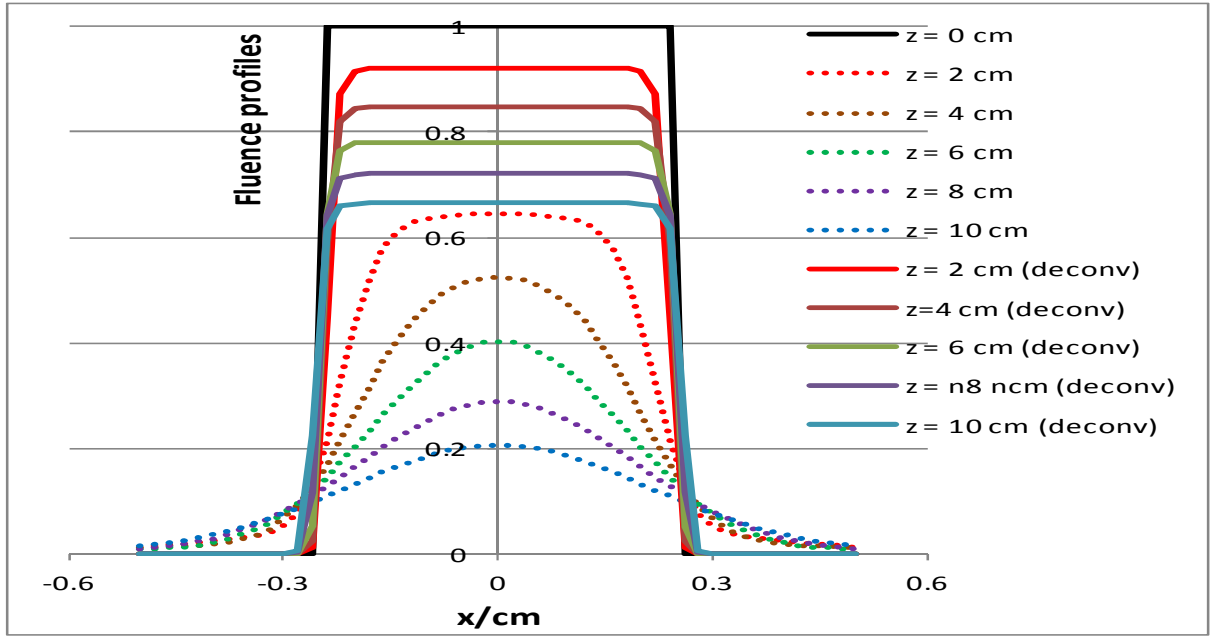

**Figure 21.** Transverse profiles at several depths of a pencil beam (field-size 4.8 x 4.8 mm$^2$) and the deconvolutions to gain the true absorption curve according to Figure 20.

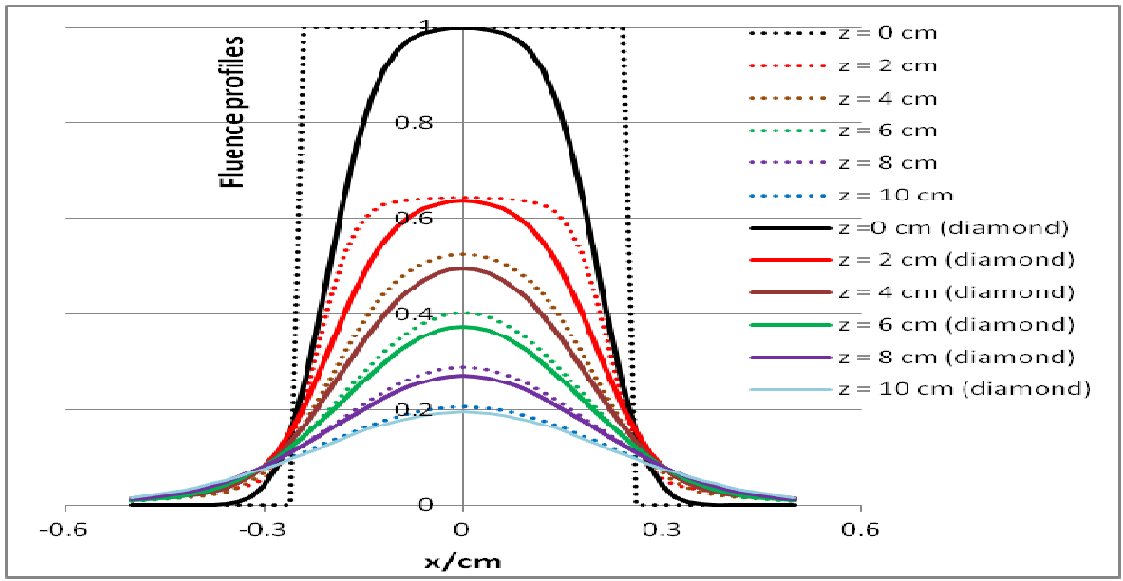

**Figure 22.** Lateral profiles according to Figure 22 (dots), the solid curves represent measurement data obtained by a diamond detector with σ = 1 mm.

This particular requirement of the previous method cannot be satisfied in the following case, where we consider a very small proton beam of interest in all scanning methods and in IMPT with additional intensity modulation. The field-size again amounts to 4.8 x 4.8 mm$^2$, which we have used due to available data of the HCL.

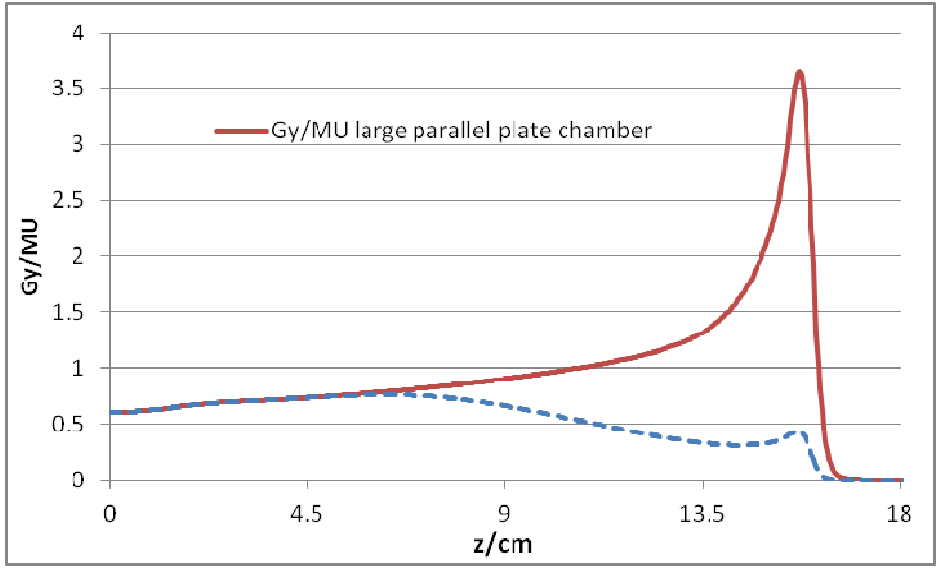

**Figure 23.** Stopping-power of 158 MeV protons (HCL) including scatter (dashes) and absorption (solid curve).

Thus it turned out by many studies that in proton therapy scanning methods have a preferred importance, since broad beams a rather difficult to handle due to the varying range of targets and range shifts resulting from heterogeneity of patient tissue. The solid curve of Figure 23 can be measured, if the proton beam is sufficiently broad (diameter > 1.5 cm). It can also be theoretically calculated by neglect of lateral scatter of by Monte-Carlo methods [8, 37, 42]. The peculiar behavior of the dashed curve results from lateral scatter, if the *rms*-values are of the order of the beam diameter or larger. Therefore the calculation of the solid curve from the dashed curve is a principal problem of absolute dosimetry in proton scanning. Figure 23 is based on measurement data and calculation; it corresponds to Figure 23, which is restricted to relative data. A proton stopping power curve consists of two parts, namely the stopping power of proton – electron interactions (main) contribution $S_{pp}$ (primary protons) and the release of secondary particles (protons, neutrons, deuterium, tritium, etc. due to nuclear interactions) $S_{sp}$ (secondary particles). The amount of secondary depends on the initial energy.

The determination of $S_{pp}$ and $S_{sp}$ has previously been carried out [8, 37]; we only point out that $S_{sp}$ amounts for 158 MeV protons to 4.4 % of $S_{pp}$. According to requirements of the Molière theory we have to represent the lateral scatter of the $S_{pp}$ protons by two Gaussian kernels, whereas for the $S_{sp}$ particles we use only one kernel due to their minor contribution. The parameters of these kernels at the depths under consideration are given in Table 8, and the overall lateral scatter has to be weighted by $S_{pp}$ and $S_{sp}$.

**Table 8.** Parameters for deconvolution of lateral scatter at z = 6 cm and at the Bragg peak.

|  |  |  | z = 6 cm |  |  | z =(Bragg peak) = 17.1 cm |  |  |
|---|---|---|---|---|---|---|---|---|
| $c_0$ | $c_1$ | $C_{sp}$ | $s_0$[cm] | $s_1$[cm] | $s_{in}$[cm] | $s_0$[cm] | $s_1$[cm] | $s_{sp}$[cm] |
| 0.91 | 0.09 | 1 | 0.105 | 0.1605 | 0.191 | 0.7443 | 0.88352 | 0.91291 |

According to Table 8 the deconvolution at the Bragg peak can only be performed by the LNS procedure because $s_1$ decreases in this region due to different ranges of scatter protons (detour factor). The consequence of this procedure (Figure 23) is that the solid curve of Figure 22 can be calculated and full agreement is obtained.

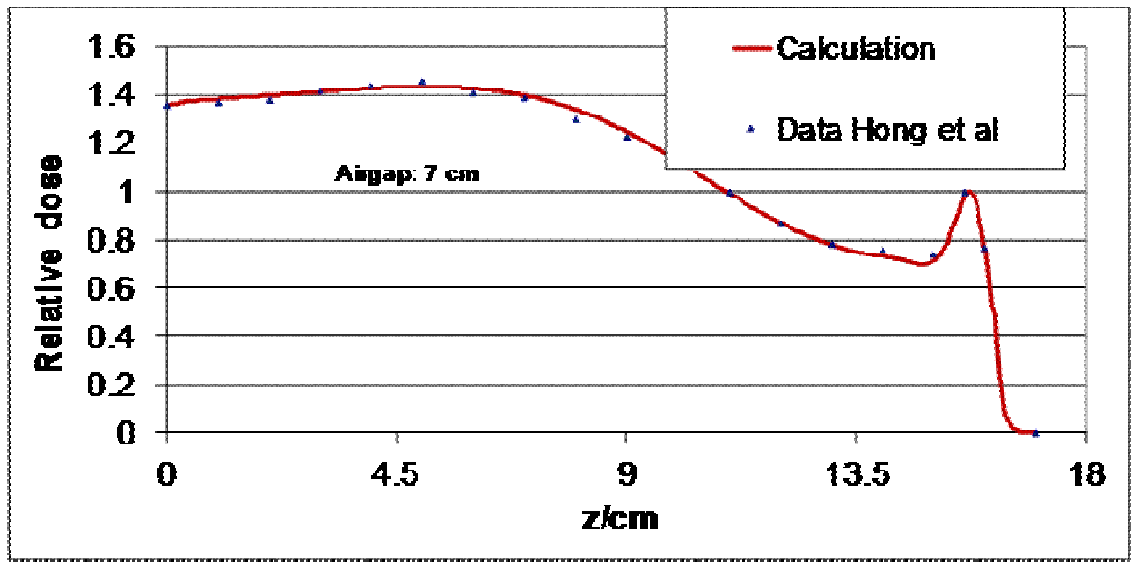

**Figure 24.** Depth dose curve of a narrow proton beam (HCL:158.6 MeV, field width of the impinging beam: 4.8 mm, measurement data in [36]).

Since the previous deconvolution method with reference to two Gaussian kernels is not applicable in the Bragg peak domain, we are not able to present a comparison. The contribution of primary protons $S_{pp}$ has been subjected to deconvolutions with N = 15 and L = 15 and of secondary particles $S_{sp}$ with N = 15 (one single Gaussian). Thus only in the initial plateau z = 6 cm the previous method would be applicable.

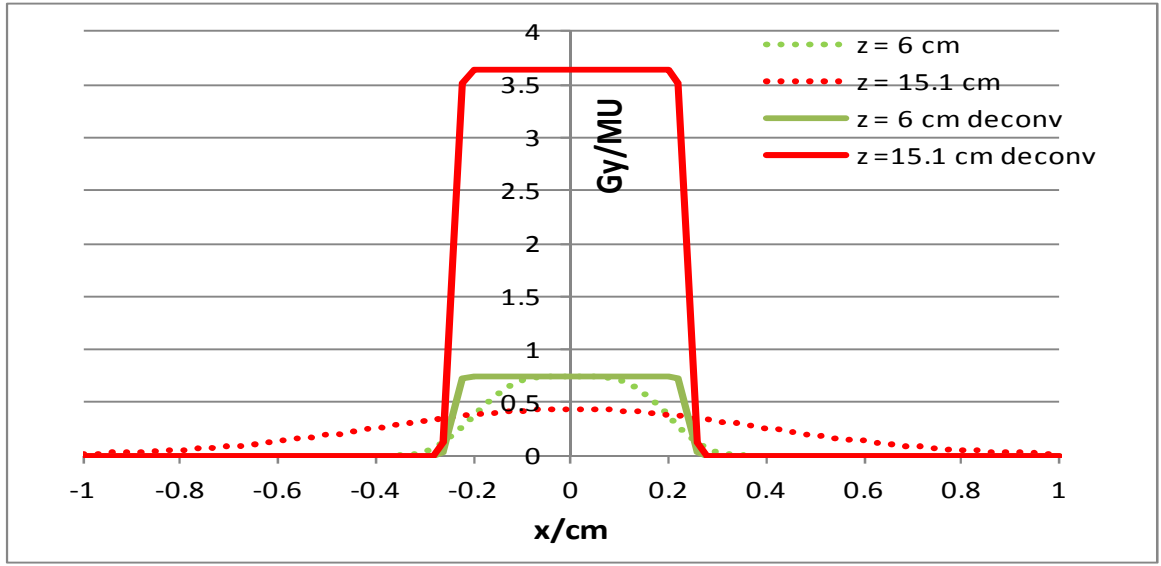

**Figure 25.** Transverse profiles of the proton scanning beam, field-size: 4.8 x 4.8 mm$^2$ including later scatter (dots) and after deconvolution (solid curves) related to Figures 23 and 24.

The importance of the deconvolution of very narrow proton beams is demonstrated by Figure 23, which is closely related to Figures 24 and 25. With regard to scanning beams in proton radiotherapy and IMPT the deconvolution can provide reliable information on the necessity of superposition of neighboring proton beamlets to avoid underdosage in a domain of interest or fluence modulation in IMPT technique.

## 3.2. Applications of generalized convolutions/deconvolutions (Fermi-Dirac statistics) to electron capture

In the following we present results of calculations for protons, He ions and carbon ions; the initial energy amounts to 400 MeV/nucleon. This appears to be a reasonable restriction with regard to therapeutic conditions. Thus Figure 26 shows that at the end of the projectile track all charged ions nearly behave in the same manner.

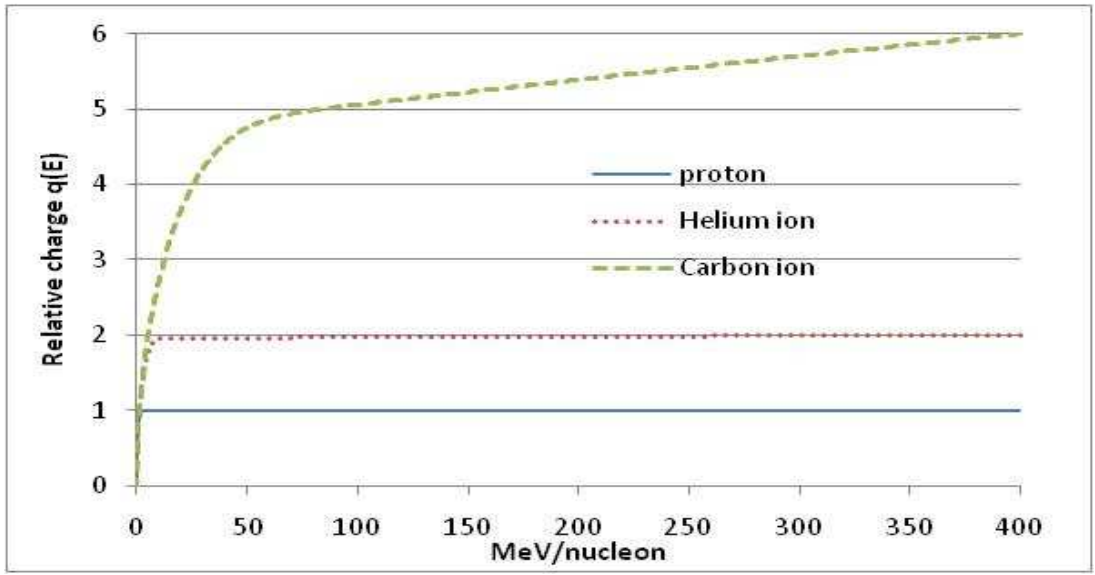

**Figure 26.** Actual charge of protons, Helium and Carbon ions in dependence of the residual energy /MeV/nucleon).

The following Figure 27 provides a more detailed behavior in the low energy domain. The residual energy per nucleon amounts to 10 MeV or smaller.

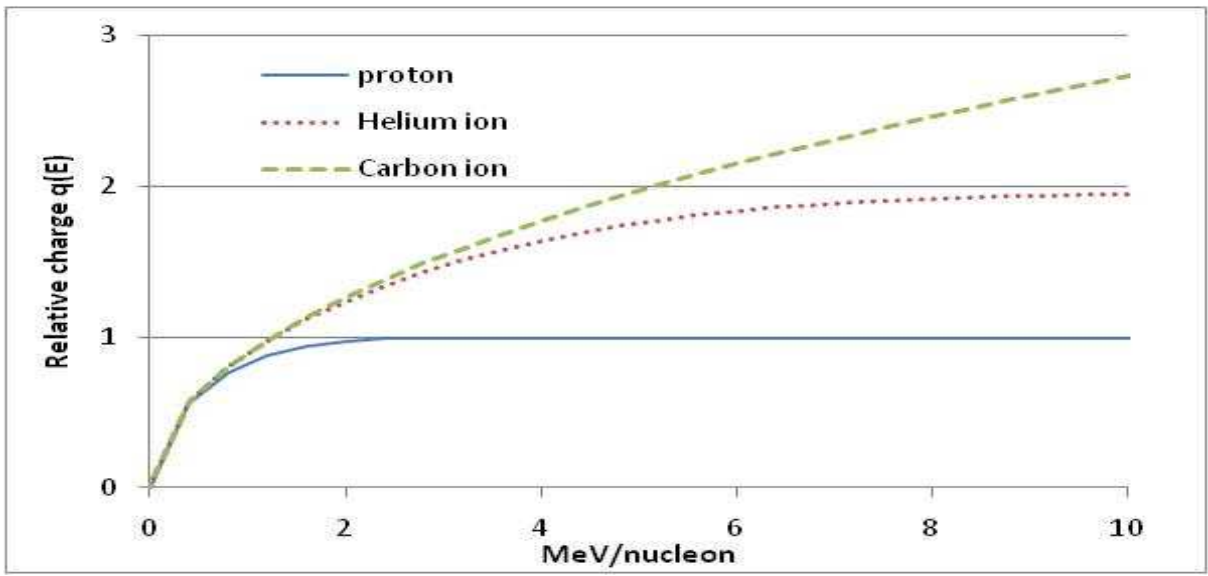

**Figure 27**. Section of the above figure for E ≤ 10 MeV.

The succeeding Figure 28 presents the decrease of the actual charge of carbon ions in dependence of the initial energy $E_0$/nucleon. Thus we can conclude that for residual energies E < 50 MeV/nucleon the behavior of the carbon ions does not depend on the initial energy $E_0$.

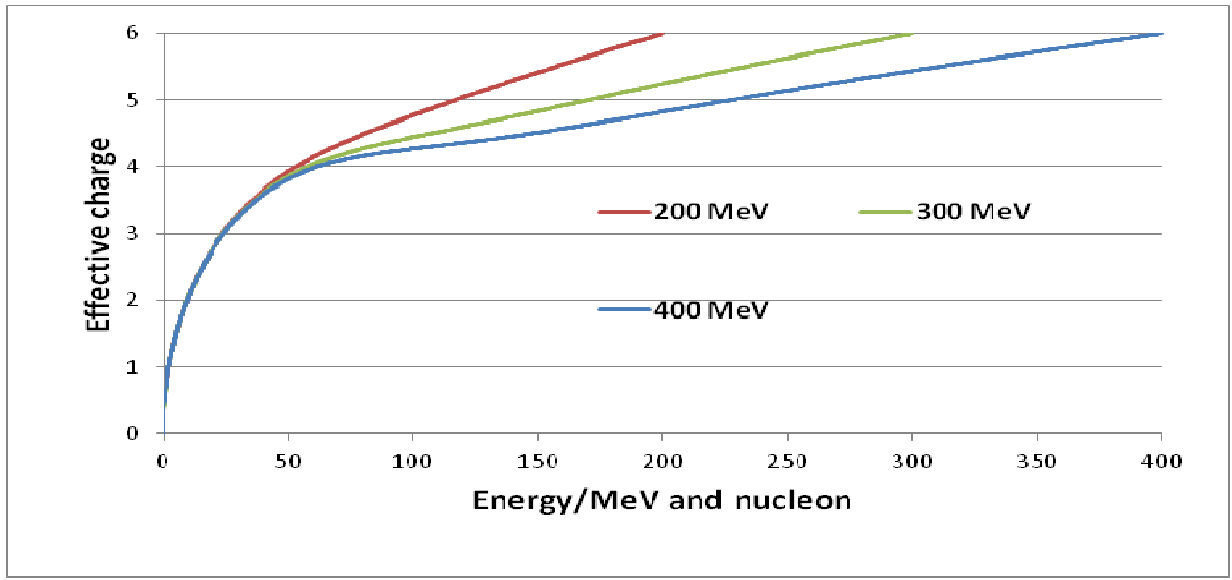

**Figure 28.** Effective charge q(E) of carbon ions in dependence of the initial energy for the cases $E_0$ = 200, 300 and 400 MeV/nucleon.

With regard to the therapeutic efficacy the behavior of the LET in the environment of the Bragg peak is very significant. For a comparison, we first regard a previous result [61, 64, 65] referring to the LET of protons. According to Figure 28 the stopping power of protons at the end track depends significantly on the initial energy $E_0$ and on the beam-line (energy spectrum at the impinging plane). The electron capture of the proton at the end track is ignored. However, the previous Figure 28 clearly shows that with regard to protons the electron capture only becomes more and more significant, when the actual proton energy is smaller than E = 2 MeV. The electron capture of protons at the end track would make the LET of protons zero independent of the initial energy.

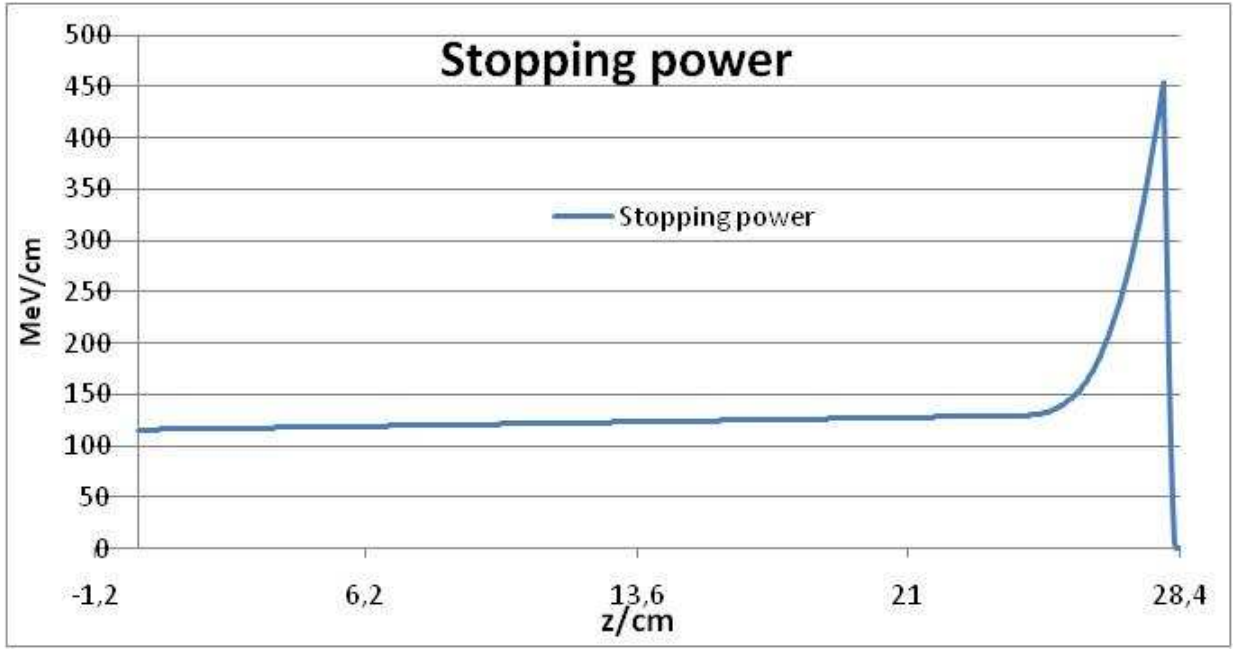

**Figure 29.** Stopping power of 400 MeV carbon ions based on the csda-approach.

The succeeding Figure 29 presents E(z) and S(z) = dE(z)/dz of protons and S(z) of carbon ions with taking account for electron capture. The initial proton energy amounts to 270 MeV, whereas the initial carbon ion energy is 400 MeV/nucleon. Most significant is the height of the Bragg peak, which is resulting from the electron capture only a factor 1.7 higher than that of protons. In both cases the csda approach is assumed. Since protons are much more influenced by energy straggling and scatter, their peak height are reduced again, whereas for carbon ions scatter and energy straggling do not play a very significant role due to the mass factor 12.

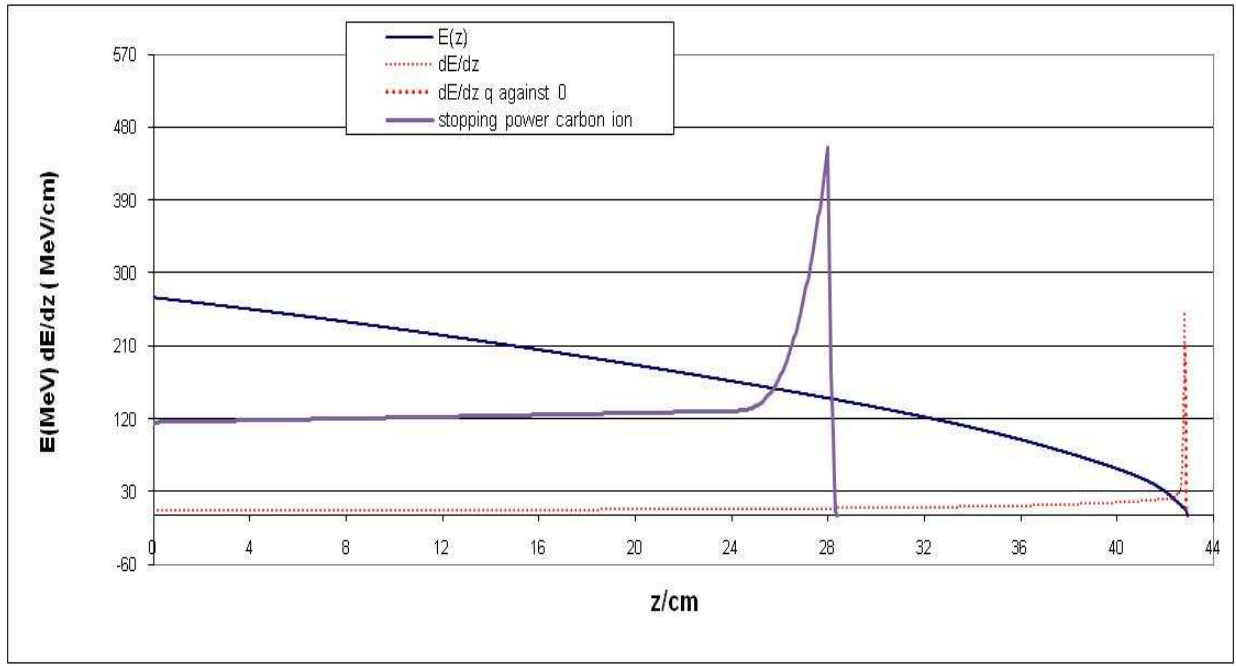

**Figure 30.** LET for mono-energetic protons (dots) and overall stopping power S(z) of carbon ions 400 MeV/nucleon.

A rigorous consideration of the LET of carbon ions is given the following Figure 31. It makes only sense to consider the total energy of 4800 MeV of the carbon ions. Due to this order of magnitude E(z) of the carbon ion has not been presented in Figure 30. Energy straggling and scatter have been ignored in Figure 31, which is justified for heavy carbons. On the other side, this figure makes also apparent the well-known disadvantage of carbon ions, namely the enormous amount of energy of carbon ions in order to reach an acceptable dose distribution in the domain of the target, where a SOBP is required. With the help of GEANT4 a real depth dose curve (HIMAC, 290 MeV/nucleon [51 – 52]) has been determined. The role of GEANT4 was only to account for the nuclear reactions, which are based in this Monte-Carlo code on an evaporation model. The electronic stopping power S(z) has been determined by the tools worked out in this communication, the electron capture effect has been accounted for. Further parameters for a calculation of S(z) have been used based on the proton calculation model [6, 62] by appropriate modifications. The Gaussian convolution kernels for energy straggling and lateral scatter have been rescaled according to the corresponding mass properties.

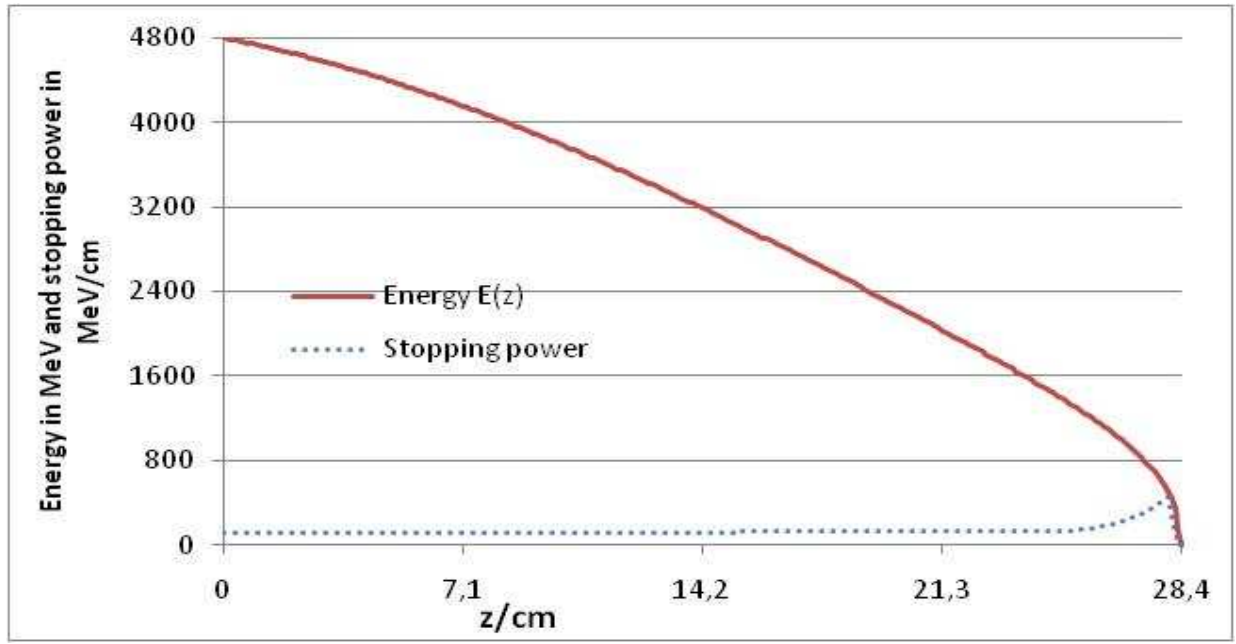

**Figure 31.** LET of carbon ions (400 MeV/nucleon).

With regard to the decrease of fluence of primary carbon ions we have derived some modifications of the corresponding decrease curves for protons. However, it appears not to be appropriate to go into further details. A further aspect is the use of the code GEANT4. Since this Monte-Carlo code represents an open programming package, some suitable additional reaction channels have been introduced.

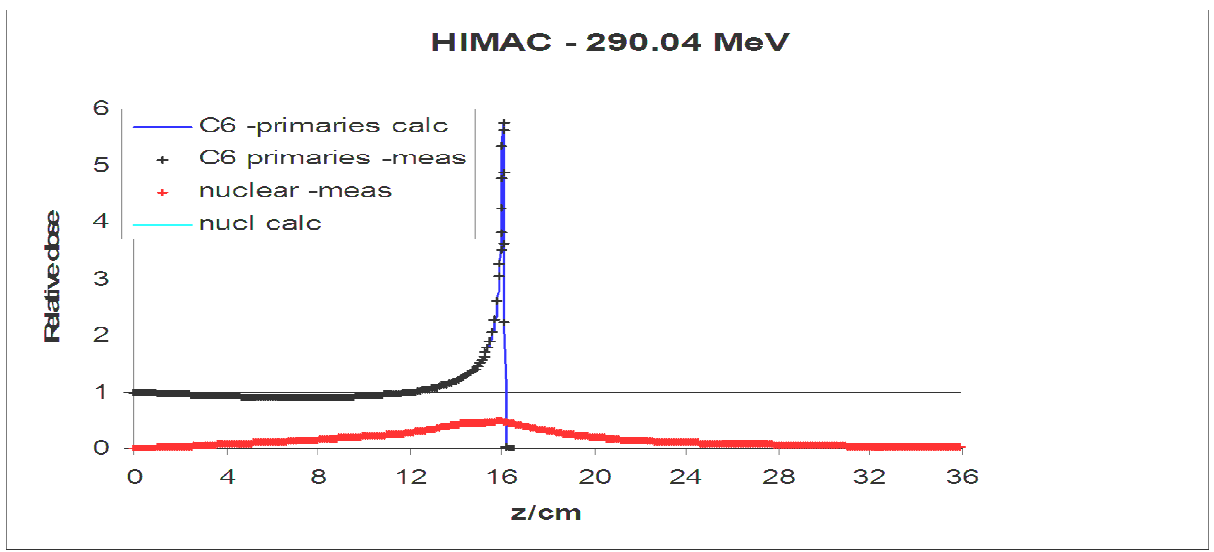

**Figure 32.** Measurement (HIMAC) and theoretical calculation of the Bragg curve of carbon ions (290 MeV/nucleon.

# 4. Discussion

The LNS-procedure is applicable with regard to the inverse problem of linear combination of Gaussian convolutions. The application of the Fermi-Dirac statistics (instead of Boltzmann) can be handled with linear combinations of shifted Gaussian convolution kernels. Thus desired back calculations can also be carried out with the LNS-procedure, e.g. the calculation of $q^2(E)$ of measured Bragg curves of heavy carbons.

## 4.1. *LNS procedure*

The main purpose of this presentation was a comparison between two different ways with respect to deconvolutions of linear combinations of Gaussian convolution kernels. As already mentioned the previous study [5] is only applicable within a more rigorous restriction with regard to the *rms*-values of the kernels: *$s_1 > s_0 \cdot 2^{0.5}$, $s_2 > s_0 \cdot 2^{0.5}$ ($s_1 \neq s_2$)*. The IFIE2 method together with the LNS solution procedure developed in this study only requires the condition: *$s_1 > s_0$, $s_2 > s_0$ ($s_1 \neq s_2$)*. If the previous procedure is applicable, then there is no principal difference to the present one with regard to accuracy and calculation speed. However, the IFIE2 method should be preferred due to its increased ability of possible applications. This is particularly true with regard to those inverse problems, where the *rms-values* do not remain constant, but can be functions of the space coordinates according to equations (100, 101), and the satisfaction of the restrictions of the previous method cannot be predicted. This fact is true for deconvolution problems of images obtained by CBCT and transverse profiles of proton Bragg curves. In particular, the inverse problem of 3D images resulting from a 360° rotation of a radiation source and the related detectors can be significantly simplified by a 3D voxel integration. An alternative way would be the deconvolution of all beamlets starting at the beam entrance and ending at the detector array. However, such a procedure consumes a lot of computation time and, by that, it is cumbersome and should be avoided. The convolution/deconvolution of boxes according to Figures 9 – 18 provide a clear indication, that the presented method shows an advantage in those cases, where discontinuities exist. The classical way is the Fourier transform together with Wiener Filters, which can lead to awkward problems at jumps of the density.

## 4.2. *Electron capture of charged particles described by generalized convolution kernels*

A further purpose of this communication was the derivation of a systematic theory of electron capture of charged particles and the role for the LET. There are purely empirical trials to include charge capture in Monte-Carlo codes. However, it appears that a profound basis for the calculation of $q^2(E)$, $E(z)$, $S(z)$ and $R_{csda}(E_0)$ depending besides the initial energy $E_0$ also on the nuclear mass number N is required to account for further influences of Bragg curves such as the density of the medium and its nuclear mass/charge $A_N$ and Z. The unmodified use of BBE leads to wrong results and the Barkas correction, which does not affect the factor $q^2$ of BBE, only works for protons or antiprotons, whereas for projectile particles like He or carbon ions this correction cannot be considered as small. The presented theory includes the Barkas effect without any correction model.

## 5. Conclusions

The property of scatter functions to account for their 2D or 3D dependence; this fact simplifies to determine the origin images by a formal way, i.e. the removal of the scatter via a calculation procedure. Scatter processes represent an inevitable property of imaging and radiation dosimetry. Besides these aspects of the inverse problem, we mention the determination of the fluence in IMRT/IMPT and refer to specific publications, where the inverse problem of Gaussian convolution plays a significant role [23, 24] and electron capture along the track of a charged particle. The discussed model cases of adjacent and nonadjacent boxes may become a significant basis for these situations. The preceding sections show that the application of the LNS method provides an attractive alternative way to solve the inverse problem (deconvolutions) of the determination of the origin image (source functions), which have been blurred by scatter of high energy photons (KV- and MV-domain). The method can be best demonstrated by model cases (phantoms). In particular, we are able to show that with regard to inverse calculations one has to be very careful in order to avoid artifacts produced by improper scatter parameters. We particular point out the problem of noise produced by certain types of detectors, which may lead to difficult decisions, whether the origin function contains real peaks or result from fluctuations of detector properties. As already pointed out the problem of noise is a typical problem in the low energy/dose domain. The deconvolution via LNS procedure acquires a particular meaning in the determination of absolute doses (monitor units/Gy) in scanning methods and IMPT of proton radiotherapy. Without profound knowledge of these parameters and further empirical experience in their handling it appears impossible to obtain reliable results of complex problems, which are confronted in CT/CBCT imaging. In order to restrict the scope of this study we have been unable to account for NMR or positron emission tomography (PET) image processing, although the latter two disciplines have become a very important tool in many other domains of medicine, which are rather different from radiology and radiotherapy, e.g. neurology, surgery and molecular image processing in pharmacology.